\documentclass[journal]{IEEEtran}
\usepackage{amsfonts}
\usepackage{mathrsfs}
\usepackage{graphicx}          % Include this line if your
\usepackage{enumerate}
\usepackage{color}
\usepackage{multicol}
\usepackage{multirow}
\usepackage{hhline}
\usepackage{setspace}
\usepackage{xcolor, colortbl}
\usepackage{amsmath}
\usepackage[font=small,labelfont=bf]{caption}
\usepackage{subfigure}
\usepackage{array}
\newcolumntype{L}[1]{>{\raggedright\let\newline\\\arraybackslash\hspace{0pt}}m{#1}}

\newcommand{\commentout}[1]{}

\newcommand{\ba}{\begin{array}}
        \newcommand{\ea}{\end{array}}
\newcommand{\bc}{\begin{center}}
        \newcommand{\ec}{\end{center}}
\newcommand{\bdm}{\begin{displaymath}}
        \newcommand{\edm}{\end{displaymath}}
\newcommand{\bds} {\begin{description}}
        \newcommand{\eds} {\end{description}}%17Apr01
\newcommand{\ben}{\begin{enumerate}}
        \newcommand{\een}{\end{enumerate}}
\newcommand{\beq}{\begin{equation}}
        \newcommand{\eeq}{\end{equation}}
\newcommand{\bfg} {\begin{figure}[ht]}
        \newcommand{\efg} {\end{figure}}%Nov 5,99
\newcommand{\bi} {\begin {itemize}}
        \newcommand{\ei} {\end {itemize}}
\newcommand{\bqn}{\begin{eqnarray}}
        \newcommand{\eqn}{\end{eqnarray}}
\newcommand{\bqs}{\begin{eqnarray}}
        \newcommand{\eqs}{\end{eqnarray}}
\newcommand{\bsl} {\begin{slide}[8.8in,6.7in]}
        \newcommand{\esl} {\end{slide}}
\newcommand{\bss} {\begin{slide*}[9.3in,6.7in]}
        \newcommand{\ess} {\end{slide*}}
\newcommand{\btb} {\begin {table}}
        \newcommand{\etb} {\end {table}}%Nov 10,99
\newcommand{\bvb}{\begin{verbatim}}
        \newcommand{\evb}{\end{verbatim}}
\newcommand{\bseq}{\begin{subequations}}
        \newcommand{\eseq}{\end{subequations}}

\newcommand{\reff}[1] {{{{\bf Figure} \ref {#1}}}}
\newcommand{\refe}[1] {{Eq(\ref {#1})}}%Nov 5

\renewcommand{\thetable}{\arabic{table}}

\begin{document}
\title{Eco-Driving at Signalized Intersections: A Multiple Signal Optimization Approach}

\author{Hao~Yang, Fawaz Almutairi, and ~Hesham~Rakha,~\IEEEmembership{Fellow,~IEEE,}
\thanks{Manuscript received...}

\thanks{Hesham Rakha is with the Virginia Tech Transportation Institute, the
Charles E. Via, Jr. Department of Civil and Environmental
Engineering, and the Bradley Department of Electrical and Computer
Engineering, Virginia Polytechnic Institute and State University, VA
24061, USA. Email: hrakha@vtti.vt.edu}

\thanks{Hao Yang is with Toyota InfoTechnology Center, CA 94043, USA. Email:
hyang@us.toyota-itc.com}

\thanks{Fawaz Almutari is with the Department of Civil and Environmental
Engineering, Virginia Polytechnic Institute and State University,
Blacksburg, VA 24061, USA. Email: fawaz@vt.edu}

\thanks{This research was partially funded by the United States Department
of Energy, Energy Efficiency and Renewable Energy - Agreement
DE-EE0008209.}}

\markboth{IEEE Intelligent Transportation Systems Transactions and
Magazine} {Yang, Almutairi, and Rakha: Eco-driving at Signalized
Intersections: A Multiple Signal Optimization Approach}

\maketitle

\begin{abstract}
Consecutive traffic signalized intersections can increase vehicle
stops, producing vehicle accelerations on arterial roads and
potentially increasing vehicle fuel consumption levels. Eco-driving
systems are one method to improve vehicle energy efficiency with the
help of vehicle connectivity. In this paper, an eco-driving system is
developed that computes a fuel-optimized vehicle trajectory while
traversing more than one signalized intersection. The system is
designed in a modular and scalable fashion allowing it to be
implemented in large networks without significantly increasing the
computational complexity. The proposed system utilizes signal
phasing and timing (SPaT) data that are communicated to connected vehicles (CVs) together with real-time vehicle dynamics to compute fuel-optimum trajectories. The proposed algorithm is incorporated in
the INTEGRATION microscopic traffic assignment and simulation
software to conduct a comprehensive sensitivity analysis of various
variables, including: system market penetration rates (MPRs), demand
levels, phase splits, offsets and traffic signal spacings on the
system performance. The analysis shows that at 100\% MPR, fuel
consumption can be reduced by as high as 13.8\%. Moreover, higher
MPRs and shorter phase lengths result in larger fuel savings.
Optimum demand levels and traffic signal spacings exist that
maximize the effectiveness of the algorithm. Furthermore, the study
demonstrates that the algorithm works less effective when the
traffic signal offset is closer to its optimal value. Finally, the
study highlights the need for further work to enhance the algorithm
to deal with over-saturated traffic conditions.
\end{abstract}


\section{Introduction}
Over the past decades, the heavy volume of passenger cars and trucks
has led to a significant increase in vehicle energy consumption and
emissions. In 2016, the transportation sector consumed 28.7\% of the
total energy consumption in the U.S., 84\% of which was consumed by
passenger cars, trucks, and buses \cite{bts2018}. The urgent need to
reduce transportation sector fuel consumption levels requires
researchers and policy makers to develop various advanced
fuel-reduction strategies. Eco-driving is one viable and
cost-effective strategy to improve fuel efficiency in the
transportation sector \cite{jollands201025}. The main premise behind
eco-driving is to provide real-time driving advice to individual
vehicles so that drivers can adjust their driving behavior or take
certain driving actions to reduce vehicle fuel consumption and
emission levels. Generally, most eco-driving strategies work by
providing real-time driving advice, such as advisory speed limits,
recommended acceleration or deceleration levels, speed alerts, etc.
To date, numerous studies have indicated that eco-driving can reduce
fuel consumption and greenhouse gas (GHG) emissions by approximately
10\% on average \cite{barkenbus2010eco}.

The major causes of high fuel consumption levels and air pollutant
emissions generated by vehicles have been widely investigated.
Frequent accelerations associated with stop-and-go waves
\cite{rakha2003comparison,barth2008real}, excessive speed (over 60
mph), slow movements on congested roads
\cite{ding2002trip,barth2008real}, and extra idling time all
dramatically increase fuel consumption and emission levels.
Consequently, it is clear that reducing speed and movement
fluctuations and reducing idling time are two critical ways to
reduce fuel consumption levels.

In general, eco-driving research can be categorized into
freeway-based and arterial-based strategies. On freeways, the
traffic stream is continuous, and vehicles are rarely affected by
traffic signals (i.e., a vehicles can travel to a particular
destination without any extra constraints, with the exception of on
and off ramps). Generally, eco-driving strategies on freeways
compute advisory speed \cite{barth2009energy} or speed limits
\cite{yang2014control,park2011predictive,park2012predictive,ahn2013ecodrive}
for drivers with the help of vehicle-to-infrastructure (V2I) or
vehicle-to-vehicle (V2V) communications, and alter driving behavior
to minimize vehicle fuel consumption levels. Wang et
al.\cite{wang2017developing} applied cooperative adaptive cruise
control (CACC) to construct vehicle platoons to achieve the goal of
minimizing platoon-wide energy consumption on uninterrupted roads.
Unlike freeways, arterial roads have traffic control devices that
routinely interrupt the traffic stream. Vehicles are forced to stop
ahead of traffic signals when encountering red indications,
producing shock waves within the traffic stream. These shock waves
in turn result in vehicle acceleration/deceleration maneuvers and
idling events, which increase vehicle fuel consumption and emission
levels. Most research efforts have focused on optimizing traffic
signal timings using traffic volumes and vehicular queue lengths
\cite{li2004signal,stevanovic2009optimizing}. Recently, by
developing connected vehicles (CVs), individual vehicles can be
controlled to minimize fuel consumption levels. CV technology
enables vehicles to exchange road traffic information, and to
communicate with traffic signal controllers to receive signal
phasing and timing (SPaT) information \cite{connected2015}, which
can be applied to estimate fuel-optimized trajectories for vehicles
traveling on arterial roads.

In the past decade, environmental CV applications have attracted
significant research interest. Most of these efforts assist drivers
in their travel along signalized intersections by providing
fuel-optimized trajectories. Mandava et al.
\cite{mandava2009arterial}, Barth, et al. \cite{barth2011dynamic},
and Xia et al. \cite{xia2013dynamic} developed an Eco-Approach and
Departure (EAD) application, which included a velocity planning
algorithm based on traffic signal information to maximize the
probability of encountering a green indication when approaching
multiple intersections. The algorithm attempted to reduce fuel
consumption levels by minimizing acceleration/deceleration durations
while avoiding complete stops. In addition, Altan et al. introduced
a Glidepath prototype to realize the EAD system in the field
\cite{altan2017glidepath}. It should be noted, however, that
decreasing the time spent in acceleration/deceleration maneuvers
does not necessarily imply reducing fuel consumption levels. To
solve this problem, optimum speed profiles are computed for probe
vehicles to reduce fuel consumption levels
\cite{malakorn2010assessment,barth2011dynamic}. Asadi and Vahidi
applied traffic signal information to estimate optimal cruise speeds
for probe vehicles to minimize the probability of stopping at
signals during red indications \cite{asadi2011predictive}. Rakha and
Kamalanathsharma constructed a dynamic programming based
fuel-optimization strategy using recursive path-finding principles,
and evaluated it with an agent-based model \cite{rakha2011eco,
kamalanathsharma2013multi, kamalanathsharma2015network}. De Nunzio
et al. used a combination of a pruning algorithm and an optimal
control approach to find the best possible green wave if the
vehicles were to receive SPaT information from multiple upcoming
intersections \cite{de2013eco}. In addition, some practical
applications, such as Green Light Optimized Speed Advisory (GLOSA)
\cite{katsaros2011performance, katsaros2011application,
seredynski2013multi, seredynski2013comparison} systems and eCoMove
\cite{ecomove2015}, were developed to estimate optimal advisory
speeds for individual vehicles proceeding through single and
multiple traffic signals to minimize delay and fuel consumption
levels. Furthermore, various smartphone applications have been
developed to provide eco-driving assistance systems for vehicles in
the vicinity of signalized intersections
\cite{koukoumidis2011signalguru,munoz2013validating}. However, the
studies above only attempt to minimize idling time and smooth
acceleration/deceleration maneuvers without considering the impact
of surrounding traffic. While in reality, the idling events are
determined not only by the SPaT information, but also by vehicle
queues at traffic signals.

To examine the impact of surrounding traffic, Qian, et al. applied
micro-simulations to evaluate the effectiveness of eco-driving with
moderate acceleration/deceleration levels while discharging from a
queue \cite{qian2011evaluating,qian2013effectiveness}. Chen
\cite{chen2013optimization} and Jin \cite{jinpower} estimated the
optimal speed profiles for individual vehicles with the
consideration of vehicle queues and road grades. Xia et al.
\cite{xia2013development} and Hao et al. \cite{hao2017dynamic}
improved the EAD system with the consideration of surrounding
traffic to estimate vehicle speed profiles. However, these studies
assumed that the queue length are given or can be detected by other
sensors. While, this assumption is not quite realistic, as the queue
length ahead of intersection changes over time, and the queue is not
easy to be detect especially at low market penetration rates of
connected vehicles. Consequently, without integrating the queue
estimation in the algorithm development, it will be very difficult
to implement the systems in the real world. Moreover, they only
focus on optimizing the speed profiles of probe vehicles upstream of
the intersection, but ignore accelerating behaviors after the signal
turns to green, which contributes to more fuel usage for vehicles
proceeding through intersections. Wang et al. developed a
cluster-wise cooperative EAD application to improve both energy
efficiency of individual vehicles and intersection throughput by
generating vehicle clusters \cite{wang2017cluster}. However, this
application requires a 100\% market penetration rate of connected
cars. In \cite{yang2017eco,ala2016modeling}, an eco-driving
algorithm at a single intersection was developed that predicts the
vehicle queue and considers downstream accelerating behaviors.
However, the algorithm only minimized fuel consumption for vehicles
proceeding through a single intersection, which restricted its
applications on arterial corridors with multiple consecutive
intersections. In \cite{he2015optimal}, a multi-stage optimal signal
control system was applied to obtain the optimal vehicle trajectory
through multiple intersections with the consideration of vehicle
queues. However, the system becomes very complex when more than two
intersections are considered.

In this paper, we extend the eco-driving algorithm proposed in
\cite{yang2017eco,ala2016modeling} to multiple signalized
intersections. We first develop the algorithm with the consideration
of both SPaT and vehicle queue information to compute fuel-efficient
vehicle trajectories in the vicinity of two or more consecutive
intersections. Subsequently, the algorithm is implemented and tested
in the INTEGRATION simulator \cite{rak2013integration}. A
comprehensive sensitivity analysis of a set of variables, including
the market penetration rates (MPRs), demand levels, phase splits,
offsets, and traffic signal spacings, is presented to quantify the
potential benefits of the proposed algorithm. This testing is
conducted on an arterial road with four signalized intersections and
a grid network composed of 16 signalized intersections. In summary,
the major contributions of the paper are: (1) Developing a vehicle
eco-drive system that considers multiple downstream traffic signals
in computing energy-efficient vehicle trajectories. The system is
proven to be more efficient than a previously developed system for
each intersection in isolation. This contribution is critical given
that in urban areas traffic signals are typically closely spaced and
thus produce interacting effects; (2) Designing a modular system
that can operate efficiently in large networks without significantly
increasing the computational complexity, thus enabling real-world
implementations; and (3) Testing the system to identify the most
efficient settings to maximize the performance of the proposed
eco-driving system and identify its limitations.

In terms of the paper layout, Section II develops an eco-driving
algorithm for multiple intersections taking vehicle queues into
consideration. Section III evaluates the algorithm with INTEGRATION
in networks with two and four intersections. Section IV includes a
comprehensive sensitivity analysis of a set of variables applied in
the algorithm. Finally, section V provides the conclusions of the
study.

\section{Eco-driving at Multiple Intersections}
In this section, the eco-driving algorithm proposed in
\cite{yang2017eco, ala2016modeling} is extended to consider multiple
intersections. The eco-driving algorithm utilizes SPaT data obtained
via V2I communications to compute fuel-optimized vehicle
trajectories in the vicinity of a signalized intersection. The
trajectory is optimized by computing an advisory speed
limit\footnote{The proposed system provides an advisory speed to the
connected vehicles. This advisory speed is overridden when the
desired headway is violated similar to on the market ACC systems. This ensures that no collisions occur and that the vehicle does not violate the car-following recommendations.} using the eco-driving algorithm, which takes the vehicle queue upstream of the intersection into consideration. However, the trajectory is only optimized for a single intersection. When traveling through multiple intersections, however, the trajectory may not minimize the vehicle's fuel consumption.

\bfg \bc
\subfigure[]{\includegraphics[width=3in,height=3in]{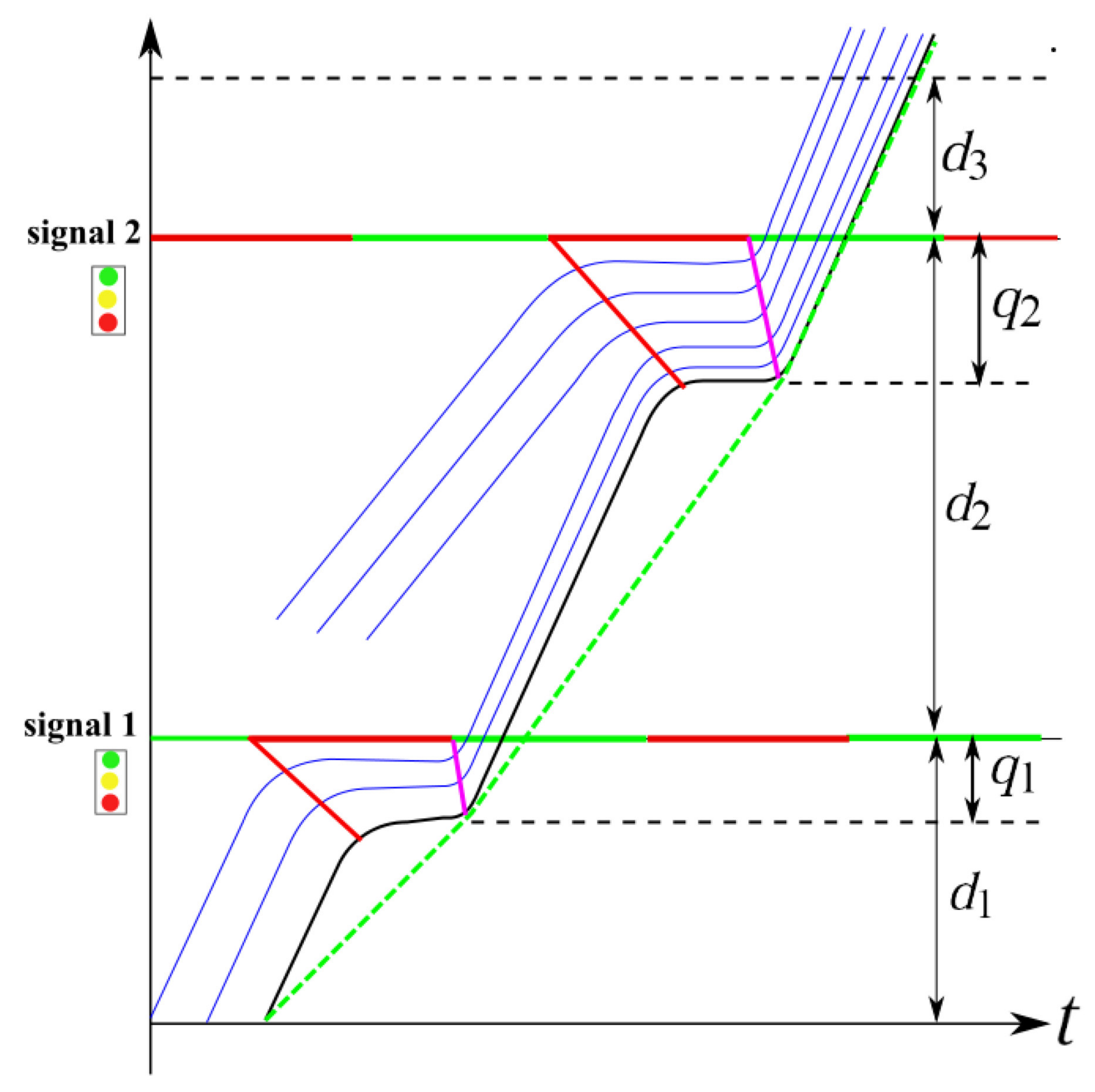}}
\subfigure[]{\includegraphics[width=3.3in,height=1.9in]{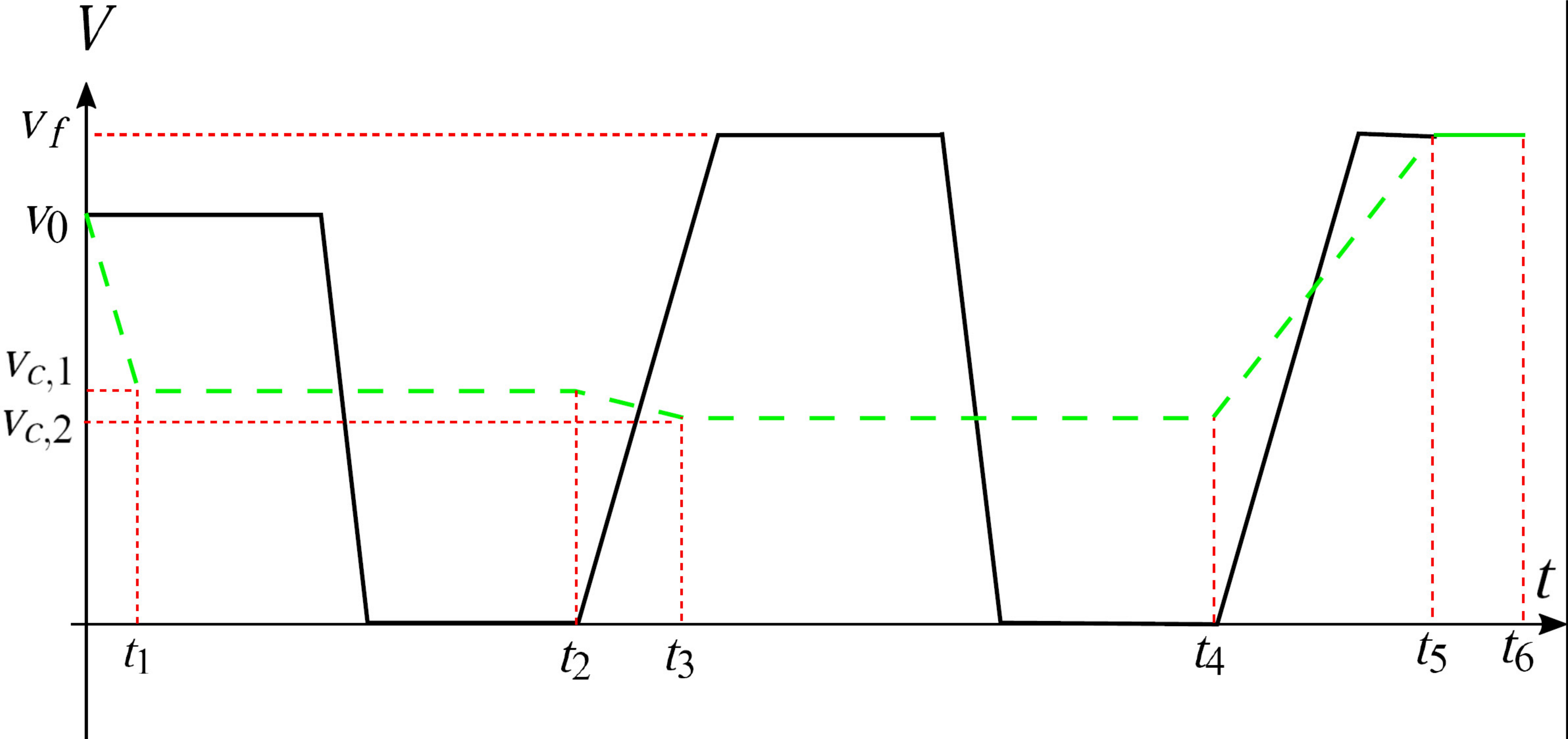}}\\
\subfigure{\includegraphics[width=2in,height=0.25in]{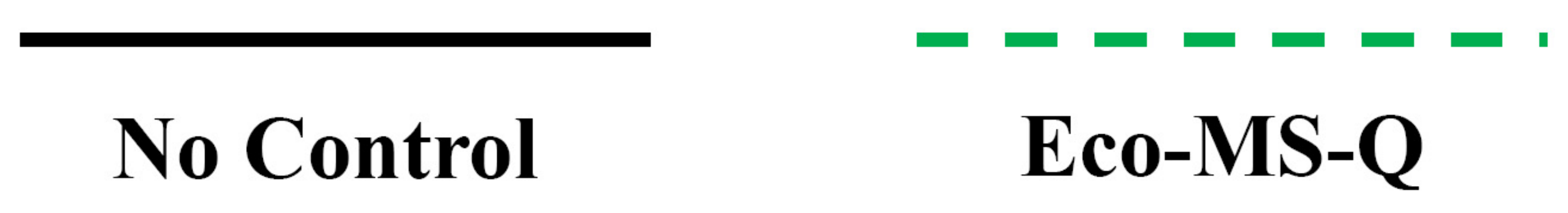}}
\caption{Dynamics of the equipped vehicle at two intersections: (a)
trajectories, (b) speed profiles.} \label{signal} \ec\efg

\reff{signal}(a) shows the trajectories of vehicles passing two
consecutive intersections. The solid black line represents the
trajectory of one vehicle experiencing two red lights without
control. The vehicle is stopped ahead of both intersections by the
red indications and the vehicle queues. Based on the work in
\cite{yang2017eco,ala2016modeling}, applying the eco-driving
algorithm for multiple intersections with the consideration of
vehicle queues (Eco-MS-Q), the vehicle cruises to each intersection
with a constant speed (see the dashed green line in
\reff{signal}(a)). \reff{signal}(b) compares the speed profiles of
the vehicle with (green line) and without (black line) control
considering both acceleration and deceleration durations. Without
control, the vehicle has to stop completely at the first
intersection. Between the two intersections, the vehicle first
accelerates to the speed limit and then decelerates to 0 again. The
stop-and-go behaviors and the long idling time waste considerable
energy. However, with control, the vehicle decelerates to a speed,
$v_{c,1}$, and cruises to the first intersection. Between the two
intersections, it decelerates or accelerates from $v_{c,1}$ to
$v_{c,2}$, and cruises to the second intersection. Here, $v_{c,1}$
and $v_{c,2}$ are the cruise speeds to the first and second
intersection, respectively. Once the queue at the second
intersection is released, the vehicle accelerates to the speed
limit. Compared to the base case without control, both the
trajectory and the speed profile with the Eco-MS-Q system enabled
are much smoother.

The objective of developing the Eco-MS-Q algorithm is to minimize
the vehicle fuel consumption level in the vicinity of multiple
intersections. In addition to the shape of the vehicle speed shown
in \reff{signal}(b), the algorithm determines the optimum upstream
acceleration/deceleration levels of the controlled trajectory in
\reff{signal}(b). The mathematical formulation of the algorithm can
be cast as

\begin{subequations}
\beq \min_{a_1, a_2, a_3} \int_0^{t_6} F(v(t)) dt, \label{obj0}\eeq
s.t. \beq v(a_1, a_2, a_3, t)=\left\{\ba{lc} v_0 + a_1 t & 0<t\leq t_1\\
v_{c,1} & t_1<t\leq t_2 \\ v_{c,1}+a_2(t-t_2) & t_2 < t\leq t_3 \\
v_{c,2} & t_3<t\leq t_4 \\ v_{c,2}+a_3(t-t_4) & t_4<t\leq t_5 \\
v_f & t_5<t\leq t_6 \ea \right.; \label{obj1}\eeq \beq
v_{c,1}=v_0+a_1 \cdot t_1; \label{obj2} \eeq \beq v_0 \cdot
t_1+\frac{1}{2} a_1 t_1^2 + v_{c,1} (t_2-t_1) = d_1 - q_1;
\label{obj3} \eeq \beq t_2=t_{g,1}+\frac{q_1}{w_1}; \label{obj4}
\eeq \beq v_{c,2}=v_{c,1}+a_2 \cdot (t_3-t_2); \label{obj5}\eeq \beq
v_{c,1}(t_3-t_2)+\frac{1}{2} a_2 (t_3-t_2)^2+v_{c,2} (t_4-t_3) =
d_2+q_1-q_2; \label{obj6}\eeq \beq t_4=t_{g,2}+\frac{q_2}{w_2};
\label{obj7}\eeq \beq v_{c,2}+a_3 (t_5-t_4)=v_f; \label{obj8}\eeq
\beq v_{c,2} (t_5 - t_4) + \frac{1}{2}a_3 (t_5-t_4)^2 + v_f
(t_6-t_5) = d_3+q_2;\label{obj9}\eeq \beq a_-^s \leq a_1 \leq a_+^s;
\label{obj10}\eeq \beq a_-^s \leq a_2 \leq a_+^s; \label{obj11}\eeq
\beq 0 \leq a_3 \leq a_+^s; \label{obj12}\eeq

Where

\begin{itemize} \setlength{\itemsep}{1pt}
\item $F(v(t))$: the vehicle fuel consumption rate at any instant $t$ computed using the Virginia Tech Comprehensive Power-based Fuel Consumption Model (VT-CPFM) \cite{rakha2011virginia} (see \refe{vtcpfm});
\item $v(t)$: the advisory speed limit for the equipped vehicle at time $t$;
\item $a_k$: the acceleration/deceleration levels for the advisory speed limit, $k=1,2,3$;
\item $v_0$: the speed of the vehicle when it enters the upstream control segment of the first intersection;
\item $v_f$: the road speed limit;
\item $d_1$: the length of the upstream control segment of the first intersection;
\item $d_2$: the distance between the two intersections;
\item $d_3$: the length of the downstream control segment of the second intersection;
\item $t_{g,1}$: the time instant that the traffic signal indication at the first signal turns green;
\item $t_{g,2}$: the time instant that the traffic signal indication at the second signal turns green;
\item $t_k$: the time instant defined in \reff{signal}(b), $k=1,2,\cdots,6$;
\item $v_{c,1}$: the cruise speed to the first intersection;
\item $v_{c,2}$: the cruise speed to the second intersection;
\item $q_1$: the queue length at the first immediate downstream intersection;
\item $q_2$: the queue length at the second immediate downstream intersection;
\item $w_1$: the queue dissipation speed at the first immediate downstream intersection;
\item $w_2$: the queue dissipation speed at the second immediate downstream intersection;
\item $a_-^s$: the deceleration level;
\item $a_+^s$: the acceleration level.
\end{itemize}
\label{obj}\end{subequations}

\begin{subequations}
$F(v(t))$ is a function of speed $v(t)$, defined by the VT-CPFM
model \cite{rakha2011virginia}, to estimate the fuel consumption
rate based on the vehicular speed and acceleration level. \beq
F(v(t), v^\prime(t))=\left\{\ba{ll}\alpha_0 +\alpha_1 P(t) +
\alpha_2 P^2(t) & P(t)\ge 0 \\ \alpha_0 & P(t)<0 \ea\right.,
\label{vtcpfm1}\eeq where, $\{\alpha_0, \alpha_1,\alpha_2\}$ are
vehicle-specific coefficients. $P(t)$ is the vehicle power at time
$t$, and is a function of the vehicle speed and acceleration. \beq
P(t)=\frac{R(t)+m\cdot
v^\prime(t)(1.04+0.0025\xi^2(t))}{3600\eta_d}\cdot v(t),
\label{vtcpfm2}\eeq \beq \ba{ll}R(t)=&\frac{\rho_a}{25.92}C_D C_h
A_f v^2(t)\\ &+9.8066 m \frac{C_r}{1000}(c_1 v(t) + c_2)+9.8066m
G(t)\ea.\label{vtcpfm3}\eeq Here, $R(t)$ is the resistance force of
the vehicle, and $\xi(t)$ is the gear ratio, and $G(t)$ is the road
grade at time $t$. $m, \rho_a, \eta_d, C_D, C_h, A_f$ represent the
vehicle mass, air density (1.2256 km/m$^3$ at sea level and a
temperature of 15C), the vehicle drag coefficient, the correction
factor for altitude, and the vehicle frontal area, respectively.
$C_r, c_1, c_2$ are rolling resistance parameters that vary as a
function of the road surface type, road condition, and vehicle tire
type. \label{vtcpfm}\end{subequations}

\refe{obj1} demonstrates that given the traffic state, including
queue lengths, the start and end times of the traffic signal
indications of the two intersections and the approaching speed of
the controlled vehicles, the speed profile varies as a function of
the acceleration/deceleration levels, $\{a_1, a_2, a_3\}$.
\refe{obj}(c-e) demonstrates that the equipped vehicle decelerates
to $v_{c,1}$ and travels through the first intersection just when the queue is released. \refe{obj}(f-h) demonstrates that the vehicle travels the second intersection when the queue is released. \refe{obj}(i-j) shows how the vehicle accelerates back to the speed limit. The Eco-MS-Q algorithm searches for the three acceleration levels that minimize the fuel consumption of the controlled vehicle over the entire control section. The flow chart of the Eco-MS-Q algorithm is illustrated in \reff{flowchart}, and the details of the algorithm, including how it is extended to $N$ consecutive intersections (labeled as $1,2,\cdots, N$ from upstream to downstream), is described below.

\ben
\item When an equipped vehicle $k$ enters the upstream control segment of
the first intersection--i.e., the distance between the vehicle and
the stop line of the intersection $1$ (the first upstream
intersection) is less than $d_1$, the Eco-MS-Q algorithm is
activated.

\item Upstream of the first intersection $1$ or the section between the
intersection $i-1$ and $i$, $i=2,3,\cdots, N$.

The algorithm estimates the optimal vehicle trajectory to proceed
through the intersections using SPaT data and vehicle queue
information \footnote{The estimation of the queue lengths,
$\{q_1,q_2\}$, and the time to release the queue is presented in
\cite{yang2017eco,ala2016modeling}.}. The algorithm categorizes the
traffic condition into three scenarios, and controls the vehicle
differently in each scenario.

\ben
\item If the equipped vehicle can proceed through its immediate downstream
intersection, $i$, at its current speed $v_0$ or at the speed limit
$v_f$, the algorithm imposes no constraints on the vehicle movement
setting the maximum speed as the road speed limit.

\item If the equipped vehicle is stopped by the immediate downstream
traffic signal, $i$, either because of its red indication or the
presence of a queue, but it can proceed through the second
intersection, $i+1$ without stopping, or $i=N$, the eco-driving
algorithm of a single intersection proposed by \cite{yang2017eco} is
applied to the equipped vehicle using the SPaT and the queue
information from intersection $i$. The optimal trajectory is
estimated for the equipped vehicle to proceed through the
intersection.

\item If the equipped vehicle is stopped by the red indication or the
queues at the two immediate downstream intersections, $i$ and $i+1$,
the optimization problem described in \refe{obj} is applied to find
the optimal trajectory for the vehicle to proceed through both
intersections. The function estimates three optimal
acceleration/deceleration levels $\{a_1^*, a_2^*, a_3^*\}$ of the
trajectory for the equipped vehicle to minimize the total fuel
consumption. \een

\item Downstream of intersection $N$, the algorithm computes the
fuel-optimum acceleration level from its current speed to the speed
limit $v_f$ over the distance $d_3$.

\item Once the equipped vehicle passes intersection $N$, and its distance
to the intersection is larger than $d_3$, the Eco-MS-Q algorithm is
deactivated. \een

\bfg \bc
\includegraphics[width=3.3in,height=4.9in]{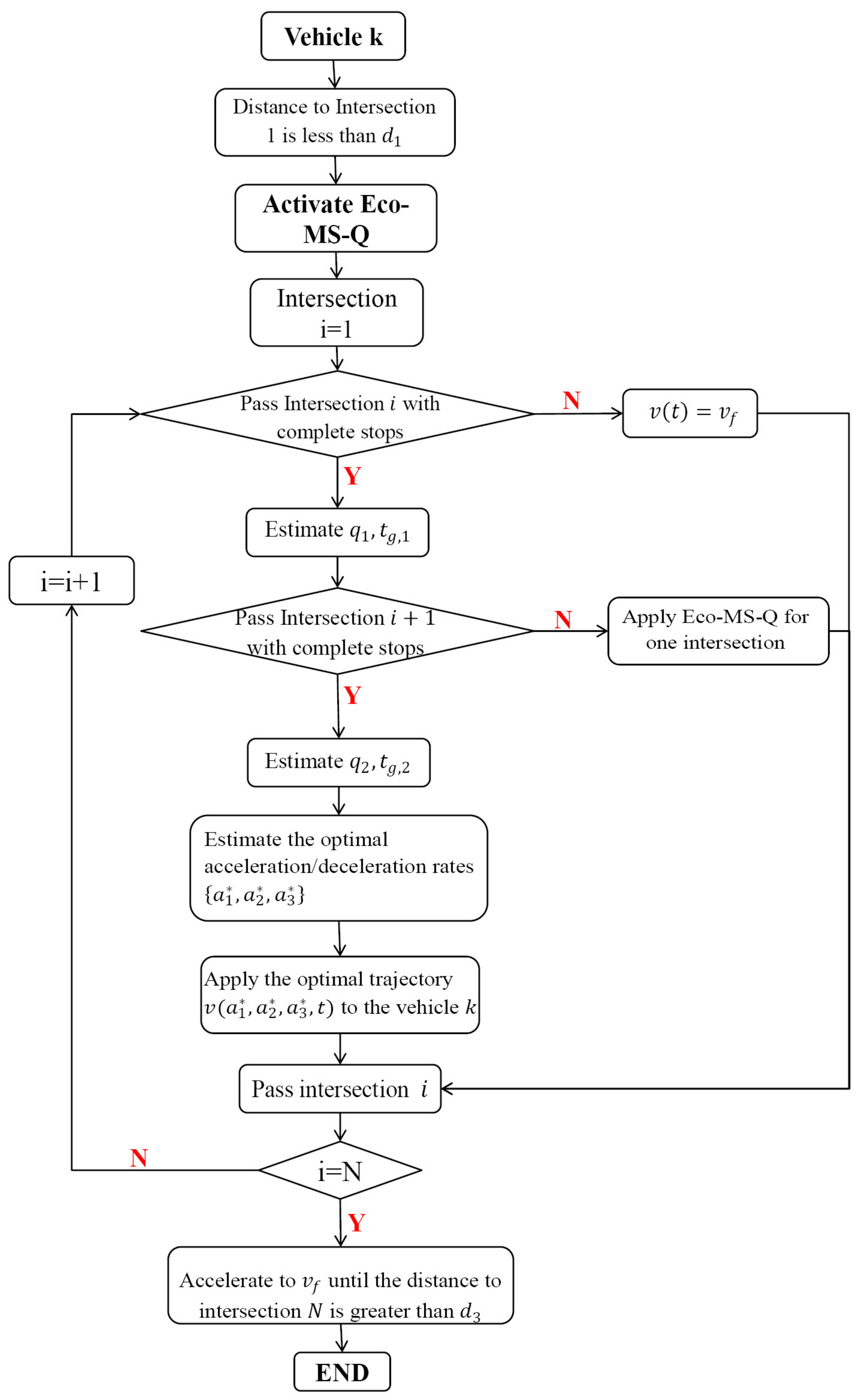}
\caption{Flow chart of the Eco-MS-Q algorithm.} \label{flowchart}
\ec\efg

The Eco-MS-Q algorithm described above applies vehicle queue
information in the estimation of the optimal trajectory. However, if
there is not sufficient information from V2I communications to
estimate vehicle queues, the algorithm can be simplified by only
using SPaT information. For that case, we developed the eco-driving
algorithm without the consideration of queue (This algorithm is
Eco-MS-O.), where the queue lengths are all assumed to be $0$, i.e.,
$q_1=q_2=0$ in \refe{obj}. In the following sections of this paper,
the two algorithms will be compared to quantify the benefits of
introducing queues in the formulation.

Note that, in the proposed Eco-MS-Q system, we only look ahead at
two downstream intersections at a time. There are two major reasons
for us to introduce this design. First, with more intersections,
there are more variables, i.e., the solution space and variables
increases significantly. The computational cost will increase
exponentially as the number of variables increases. Consequently, it
will not be easy to look ahead at too many intersections at the same
time. Second, more importantly, the search of the optimal trajectory
for the controlled vehicle relies on the prediction of the vehicle
queue length ahead of the intersections, and the queue length of
each intersection is predicted at the time when the vehicle arrives
at the intersection. With multiple intersections, the queue length
estimation of the further downstream intersections will be less
accurate and thus may produce erroneous recommendations.  Hence, the
trajectory estimation will be less efficient or even meaningless to
minimize the fuel consumption, which makes the consideration of more
intersections futile.

Note that the proposed system only provided advisory speed limits to connected vehicles. In the implementation of the porposed system, a connected vehicle first determine its desired speed using a car-following model. In this paper, the the INTEGRATION microscopic traffic simulator \cite{rak2013integration} is applied, and the embedded the Rakha-Pasumarthy-Adjerid (RPA) model \cite{sangster2014enhancing} is implemented to model car-following behaviors. The model estimates the speed of each vehicle based on the location and speed information of itself and its leader. It also has a collision avoidance component to guarantee collision-free travel. Once the speed of the vehicle is determined by the car-following model, the advisory speed limit from the proposed system is applied. If the speed of the vehicle is higher than the limit, the vehicle will further decelerate to the limit. If the speed is lower, the limit will not be applied. This advisory limit will ensure the driving safety of the connected vehicles.

\section{Evaluation of the Eco-MS-Q Algorithm}
This section implements and evaluates the benefits of the proposed
Eco-MS-Q algorithm within INTEGRATION. The simulator is capable of
modeling the traffic signal control system and the movements of
individual vehicles. In this section, the proposed algorithm is
implemented on networks with 2, 4, and 16 intersections to test its
impacts on individual vehicle dynamics, to estimate its benefits on
vehicle energy, and to check its feasibility on larger networks.

\subsection{Impact on Equipped Vehicles}
As a starting point, a simple network of two intersections defined
in \reff{simsig} is simulated. The simulation is conducted for
one-way movement where vehicles are loaded from one origin to one
destination only. In the experiment, the speed limits are $80$
km/h, and the road saturation flow rates are set at $q_c = 1600$
veh/hr/lane (veh/h/lane) for all links. In addition, we assume that there
is a reliable and perfect vehicle-to-signal communications with DSRC
devices, and it does not have any packet loss and the communication
latency is negligible.

\bfg \bc
\includegraphics[width=3.3in,height=1in]{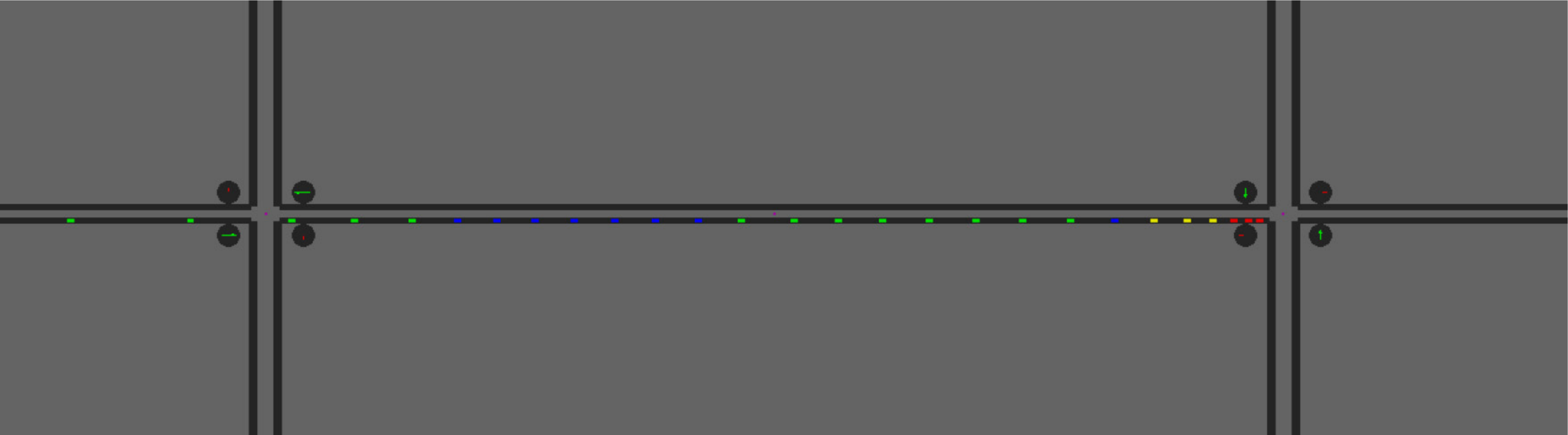}
\caption{Configuration of two consecutive intersections.}
\label{simsig} \ec\efg

Assume that the distance between the two intersections in
\reff{simsig} is $d_2=1000$ meters, the upstream control segment
length of the first intersection is $d_1=500$ meters, and the
downstream control segment length of the second intersection is
$d_3=200$ meters. \footnote{As the transmission range of DSRC can be
up to 1000 meters \cite{kenney2011dedicated}. In this paper, we
assume that the transmission range is long enough ($>$500 meter) to
cover vehicles within both the upstream and downstream control
segments to build reliable communications with the signal.} The
number of lanes of all links is set as one. In the first simulation,
vehicles are only loaded from east (left end) to west (right end) at
the rate of $600$ veh/h/lane, and 10\% of them are equipped with the
Eco-MS-Q algorithm (with the consideration of queues). For the SPaT
information, the cycle lengths of both signals are $120$ seconds,
and the durations of the green, amber, and all red indicators of the
through traffic for the first and second signals are all $61$, $4$
and $2$ seconds, respectively. The offset of the second signal with
respect to the first one is $75$ seconds.\footnote{The optimal
offset of the second signal is $45$ seconds. Here we introduce a
sub-optimal offset to evaluate the performance of the Eco-MS-Q
algorithm when the equipped vehicles experience two stops. The
75-second offset gives a high probability for us to observe two
stops for one equipped vehicle.} The equipped vehicles receive
advisory speed limits from the algorithm, which is updated every
second.

\reff{mscomp}(a) compares the trajectories of one equipped vehicle
before and after applying the Eco-MS-Q algorithm. Compared to the
trajectory without control, the trajectory (location between $500$
meters and $2200$ meters) is much smoother when the algorithm is
applied. \reff{mscomp}(b) compares the speed profiles of the vehicle
before and after applying the algorithm. Similar to
\reff{signal}(b), the equipped vehicle slows down and cruises to the
first intersection at a lower speed, and passes the intersection
without stopping. Between the first and second intersections, it
accelerates to another cruise at a moderate acceleration level and
passes the second intersection without stopping. The standard
deviations of the speed profiles before and after applying the
algorithm are $34.2$ km/h and $22.1$ km/h, respectively, i.e., the
speed oscillation is reduced by as high as 30\%. In addition, the
fuel consumption levels before and after applying the algorithm are
$0.146$ liter/km and $0.113$ liter/km, respectively, demonstrating
that the algorithm reduces vehicle fuel consumption by approximately
$22.5\%$.

Note that in \reff{mscomp}(b), after applying the algorithm, the
equipped vehicle's speed still drops ahead of the traffic signals
and fluctuates significantly, even though the vehicle does not come
to a complete stop. This fluctuation is caused by the inaccurate
estimation of the vehicle queue lengths, $q_1$ and $q_2$, and the
queue dissipation speeds, $w_1$ and $w_2$. In the Eco-MS-Q
algorithm, these variables are estimated utilizing a kinematic wave
model based on road properties, including: the road saturation flow
rate, jam density, and critical density \cite{yang2017eco}. However,
in the microscopic simulations, due to the randomness of vehicle
dynamics, the estimation cannot be accurate. Hence, the advisory
speed limits calculated by the Eco-MS-Q algorithm cannot perfectly
smooth the movements of equipped vehicles, and oscillations occur
when they travel through the intersections. In the future, with the
help of vehicle-to-vehicle communications, queue lengths and queue
dissipation speeds can be monitored in real-time to improve the
estimation of the advisory speed limits and reduce the vehicle
oscillations.

\bfg \bc
\subfigure[Trajectories]{\includegraphics[width=3in,height=2.2in]{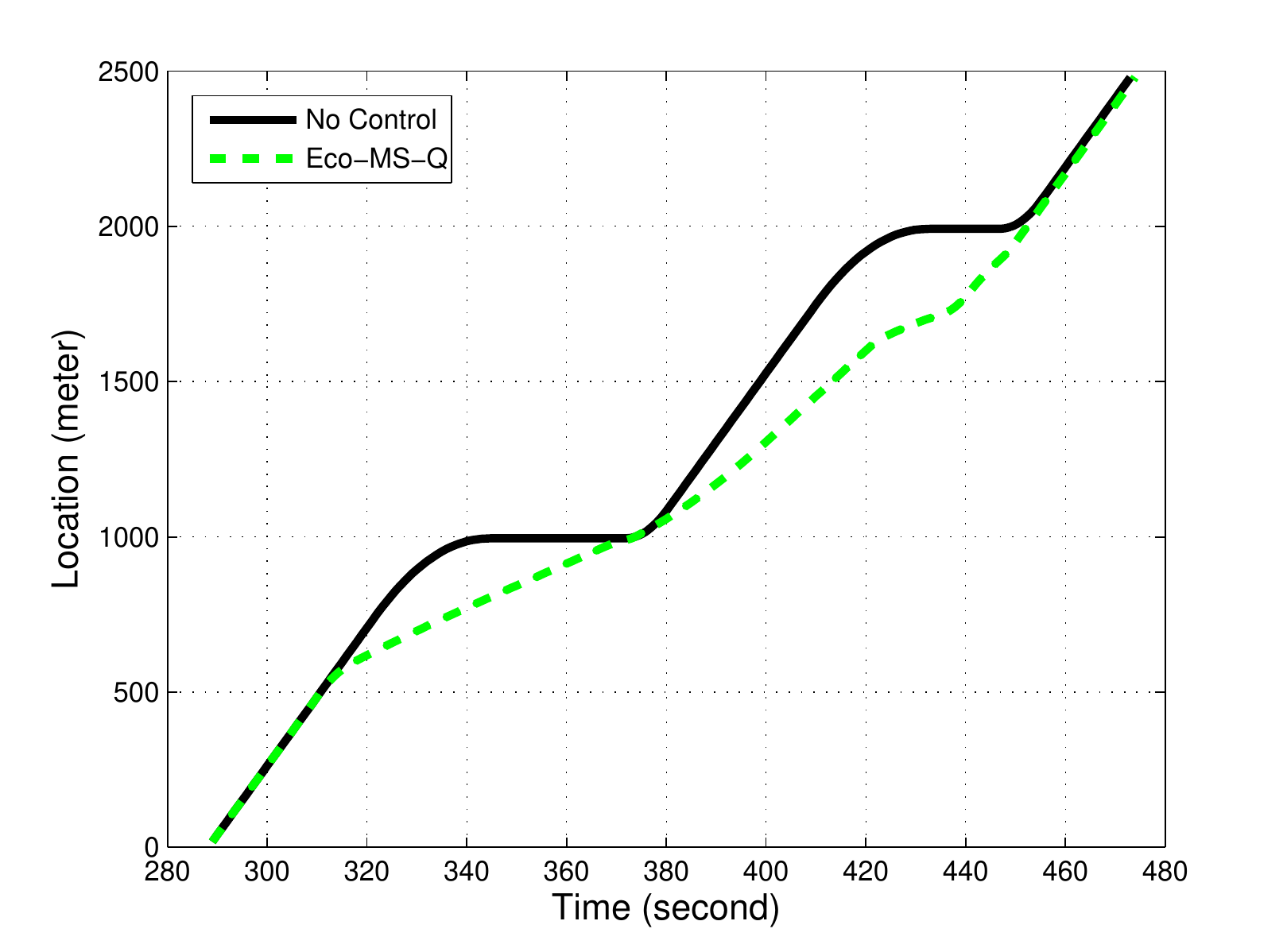}}
\subfigure[Speed
profiles]{\includegraphics[width=3in,height=2.2in]{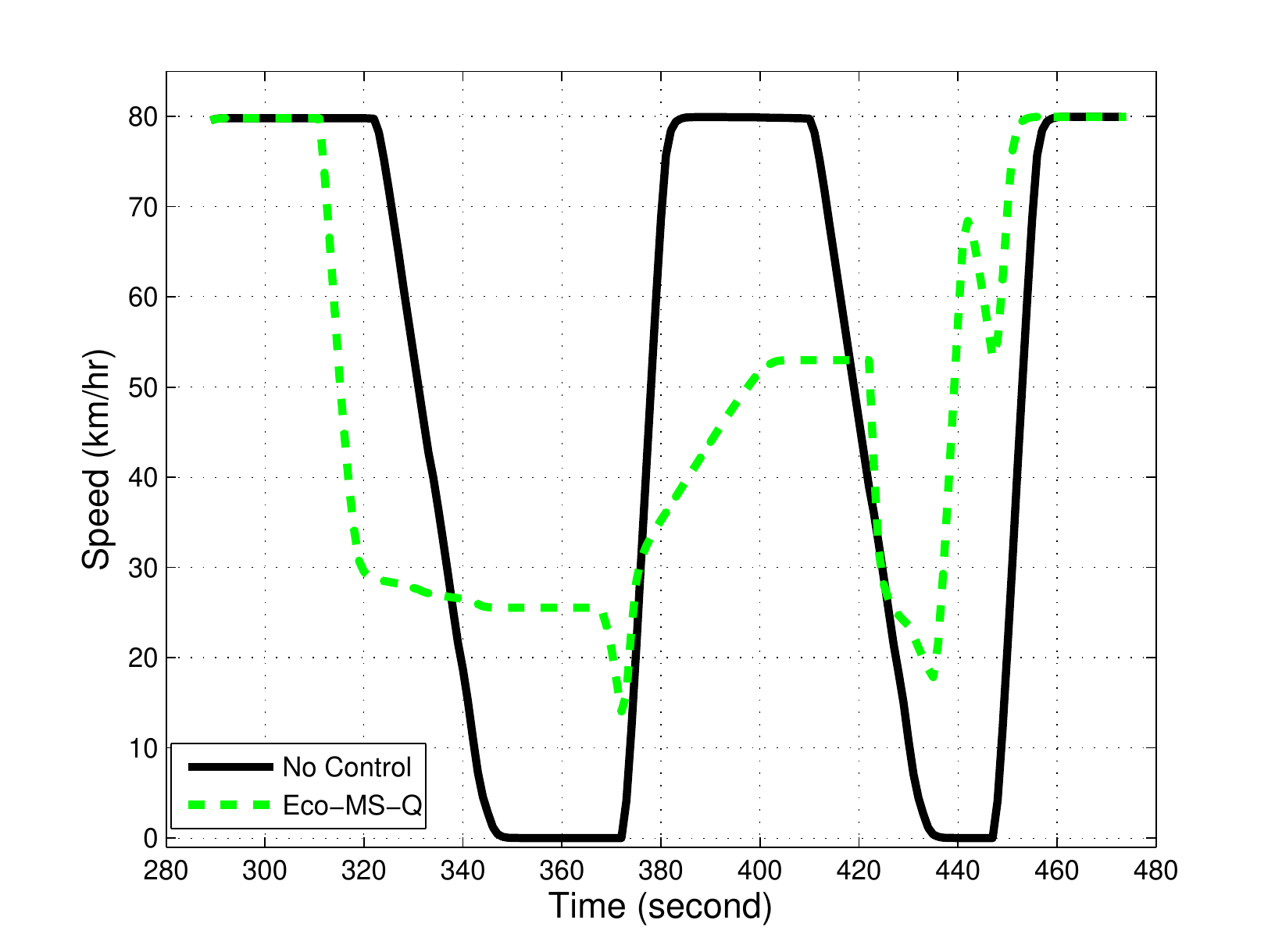}}
\caption{Comparison of vehicle movements before and after applying
the Eco-MS-Q algorithm.} \label{mscomp} \ec\efg

\subsection{Testing of Eco-MS-Q Algorithm for Two Signalized Intersections}
In this section, we evaluate the benefit of the proposed algorithm
on the network-wide fuel consumption levels, and compare the
algorithms with and without the consideration of vehicle queues
(Eco-MS-Q and Eco-MS-O) for different MPRs of equipped vehicles. In
addition, the algorithms are compared with those proposed in
\cite{yang2017eco, ala2016modeling} for independent intersections
(Eco-1S-Q and Eco-1S-O).

The network settings in section III.A are also applied in this
section. For the multiple intersection control algorithms (Eco-MS-Q
and Eco-MS-O), the equipped vehicles are under control once they are
within $500$ meters ahead of the first intersection and within $200$
meters after the second intersection for the two Eco-MS algorithms.
For the single intersection control algorithms (Eco-1S-Q and
Eco-1S-O), the equipped vehicles are under control once they are
within $500$ meters before each intersection and $200$ meters after
each intersection. This simulation was done with a single-lane
network, which prevents lane changing or vehicle over-taking
maneuvers. The demand for the network is still $600$ veh/h/lane, i.e.,
$1200$ veh/h for the two lanes, and the SPaT plans of the two signals
in section 3.1 are also applied. The offset of the second traffic
signal is set as $0$ seconds. To better evaluate the system
performance, the algorithms are tested for different MPRs; only a
portion of the vehicles are equipped, while the rest drive normally
using standard car-following models.

\bfg \bc
\includegraphics[width=3in,height=2.2in]{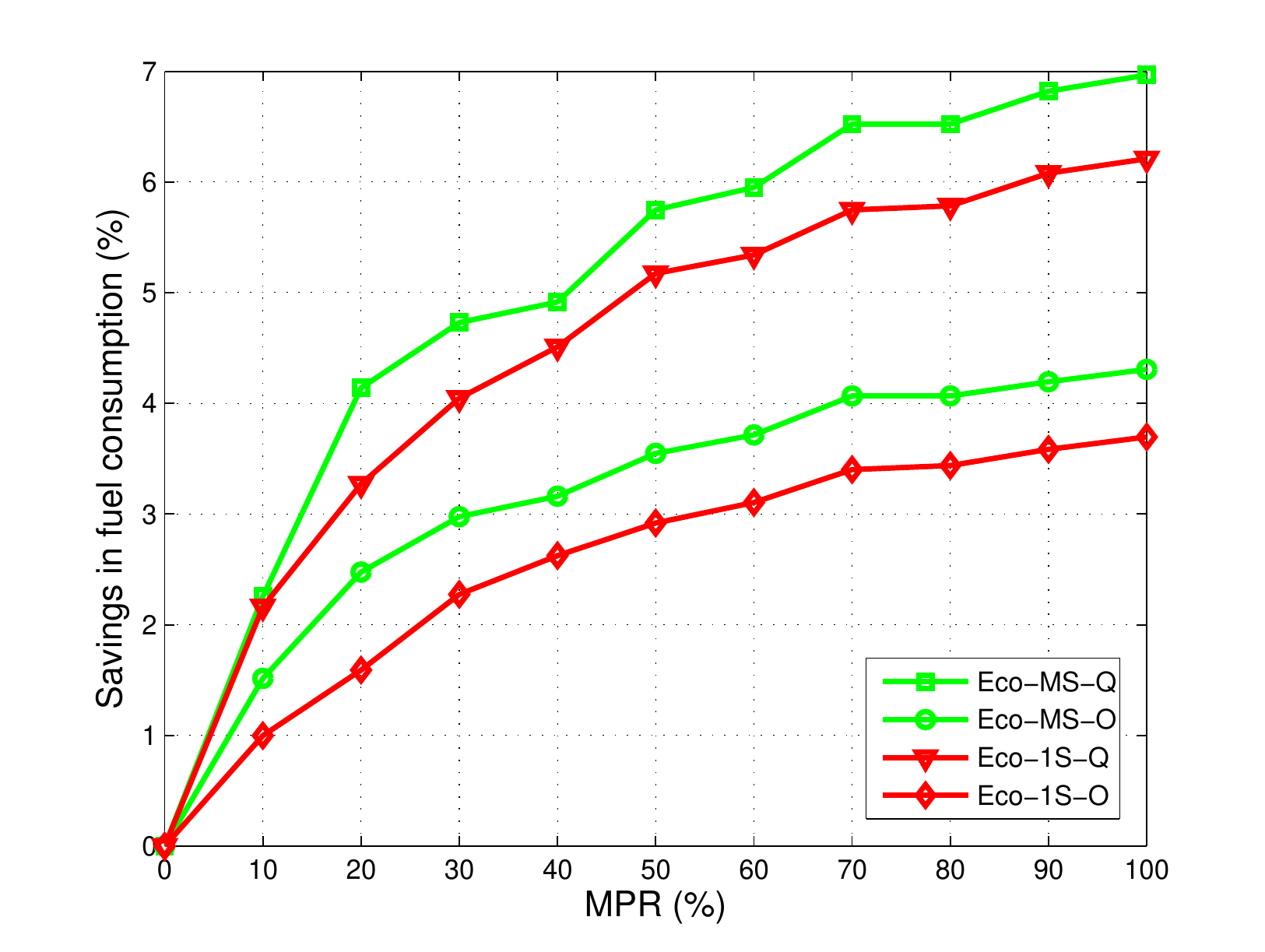}
\caption{Savings in fuel consumption for the single-lane network for
different MPRs.} \label{mpr1lan} \ec\efg

\reff{mpr1lan} demonstrates the overall network-wide energy savings
of the Eco-1S and Eco-MS algorithms considering different MPRs. The
figure illustrates that higher MPRs lead to greater savings in all
control systems. At 100\% MPR, the fuel consumption is reduced by
approximately 7\% with Eco-MS-Q and 4.2\% with Eco-1S-Q. In the
simulations, the movements of the equipped vehicles are smoothed by
the proposed algorithm, and at the same time due to the
car-following behavior, the trajectories of some non-equipped
vehicles are also smoothed, which further reduces the network-wide
fuel consumption levels. \reff{mpr1lan} also demonstrates that even
without the consideration of vehicle queues, the Eco-MS-O and
Eco-1S-O algorithms can still produce larger fuel savings as MPRs
increase. However, the savings are smaller than those that take
queues into consideration. Specifically, without considering queues,
the fuel consumption rate is reduced from 7\% to 6.1\% for the
multiple intersection control and from 4.2\% to 3.9\% for the single
intersection control.

In the simulation of the single-lane intersections above, lane
changing and over-taking behaviors are not possible; while in
reality, links with two or more lanes are common. Accordingly, the
impacts of lane changing and vehicle over-passing need to be
considered. In the second example, the same settings used in the
previous simulation are applied to the same network considering
two-lane roadways. \reff{mpr2lan} compares the fuel consumption
savings of Eco-1S-Q, Eco-1S-O, Eco-MS-Q, and Eco-MS-Q algorithms
under different MPRs. Unlike the single-lane network, the savings in
fuel consumption are not always observed in the two-lane scenarios,
especially when the MPR is less than 30\%. When MPRs are less than
30\%, all algorithms increase the overall fuel consumption levels.
The negative impact of the lower MPRs is a result of the lane
changing and passing of non-equipped vehicles. As the algorithms
only control the equipped vehicles, which are traveling at lower
speeds compared to non-equipped vehicles, larger gaps will be
generated ahead of them. The non-equipped vehicles, traveling at
higher speeds, then have a greater likelihood of changing lanes and
cutting into the gaps ahead of the equipped vehicles, increasing
traffic stream speed oscillations and thus fuel consumption levels.
With 30\% MPRs and above\footnote{Here, 30\% MPR is just an observed
threshold for the defined scenario. The value may change based on
demand levels, number of lanes, road speed limit, etc.}, the number
of equipped vehicles increases, making it increasingly possible for
equipped vehicles to travel side-by-side for the length of the link,
preventing lane changing and passing maneuvers, and increasing fuel
consumption savings. All algorithms provide positive savings with
MPRs higher than 30\%. At 100\% MPR, the network-wide fuel
consumption is reduced by approximately 6.5\% for the Eco-MS-Q,
5.8\% for Eco-MS-O, 4.2\% for Eco-1S-Q and, and 3.2\% for Eco-1S-O.
Similar to the single-lane example, the algorithms taking queue
effects into consideration always result in better performance for
the network with two-lane links.

\bfg \bc
\includegraphics[width=3in,height=2.2in]{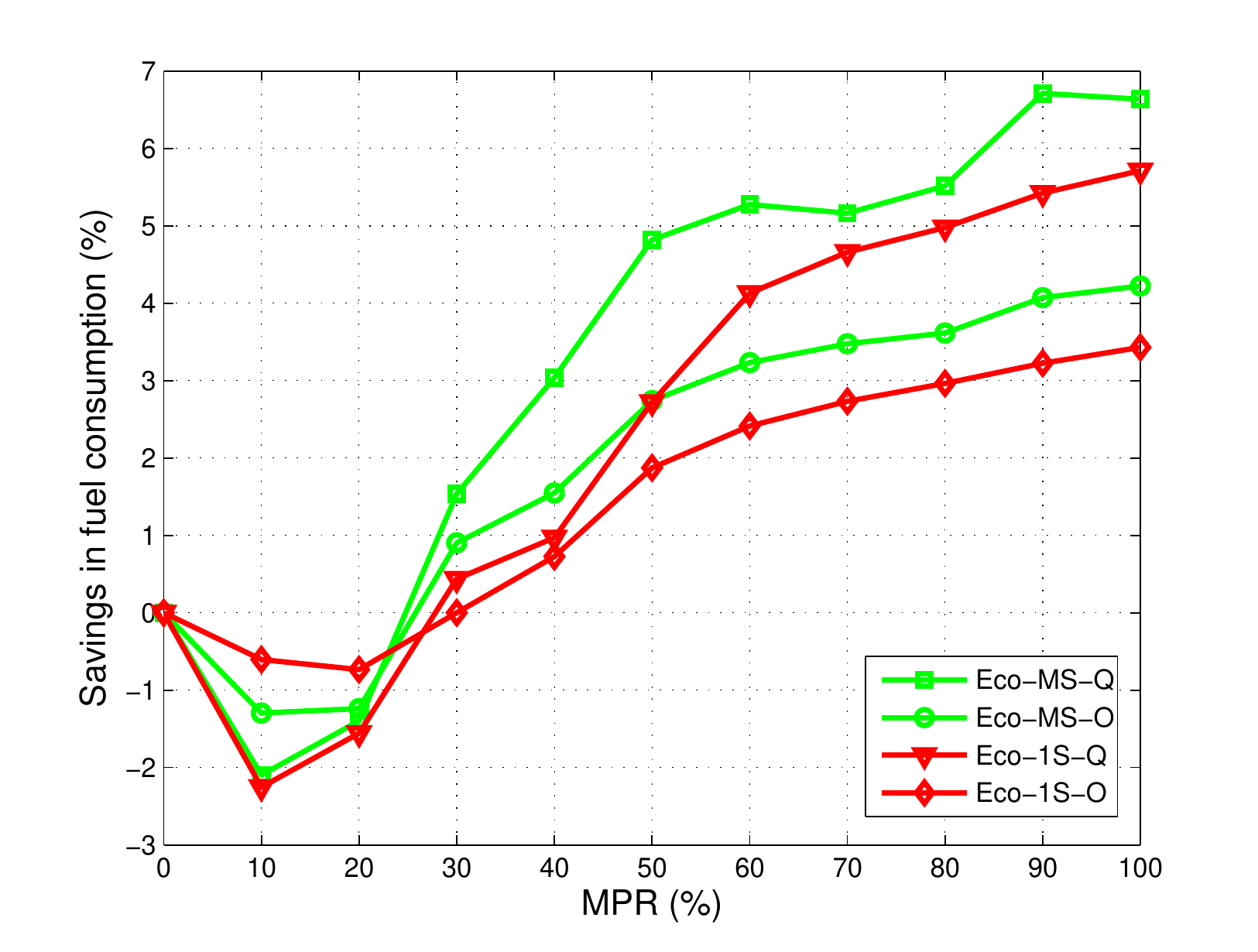}
\caption{Savings in fuel consumption for the two-lane network for different MPRs.} \label{mpr2lan} \ec\efg

\subsection{Testing Eco-MS-Q Algorithm on Arterial Network}
In addition to two-intersection networks, the Eco-MS-Q algorithm is
also applied to a larger network with more than two intersections.
The network in \reff{ms4sig}(a), which has four consecutive
intersections, is simulated considering a traffic signal spacing of
$600$ meters. The demand from the west to east is constant at $600$
veh/h. For the SPaT plan, the cycle lengths and phase splits of all
intersections are 120 seconds and 50\%, respectively, and the
offsets of all signals are set as $0$.

\reff{ms4sig}(b) illustrates the fuel consumption savings from the
Eco-1S-Q and Eco-MS-Q algorithms for different MPRs in single-lane
and two-lane networks. For the single-lane network, both algorithms
have positive benefits on network-wide fuel consumption for
different MPRs, and higher MPRs result in greater savings for both
algorithms. At 100\% MPR, fuel consumption is reduced by 7.7\% with
Eco-MS-Q, and by 6.2\% with the Eco-1S-Q algorithm. Similar to the
results in section III.B, the savings come from both equipped
vehicles with the optimal trajectories and non-equipped vehicles
following them. For the two-lane network, negative effects of the
algorithms are still observed when the MPR is low ($<15\%$ for
Eco-1S-Q and $<25\%$ for Eco-MS-Q). When the MPR is larger than
30\%, positive savings in fuel consumption can be obtained for both
algorithms, and the savings from the multiple intersection control
is higher than the single intersection control (even though the
difference is not large). At 100\% MPR, both algorithms can reduce
the fuel consumption by as high as 4.8\%. The two simulations also
demonstrate the effectiveness of the proposed algorithm on networks
with multiple intersections.

\bfg \bc
\subfigure[]{\includegraphics[width=3in,height=0.5in]{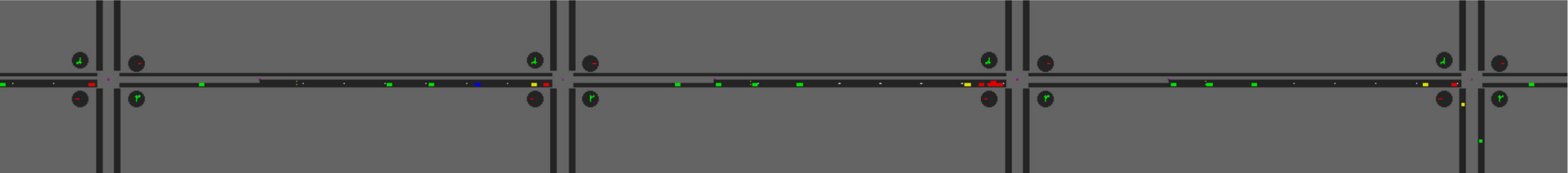}}
\subfigure[]{\includegraphics[width=3in,height=2.2in]{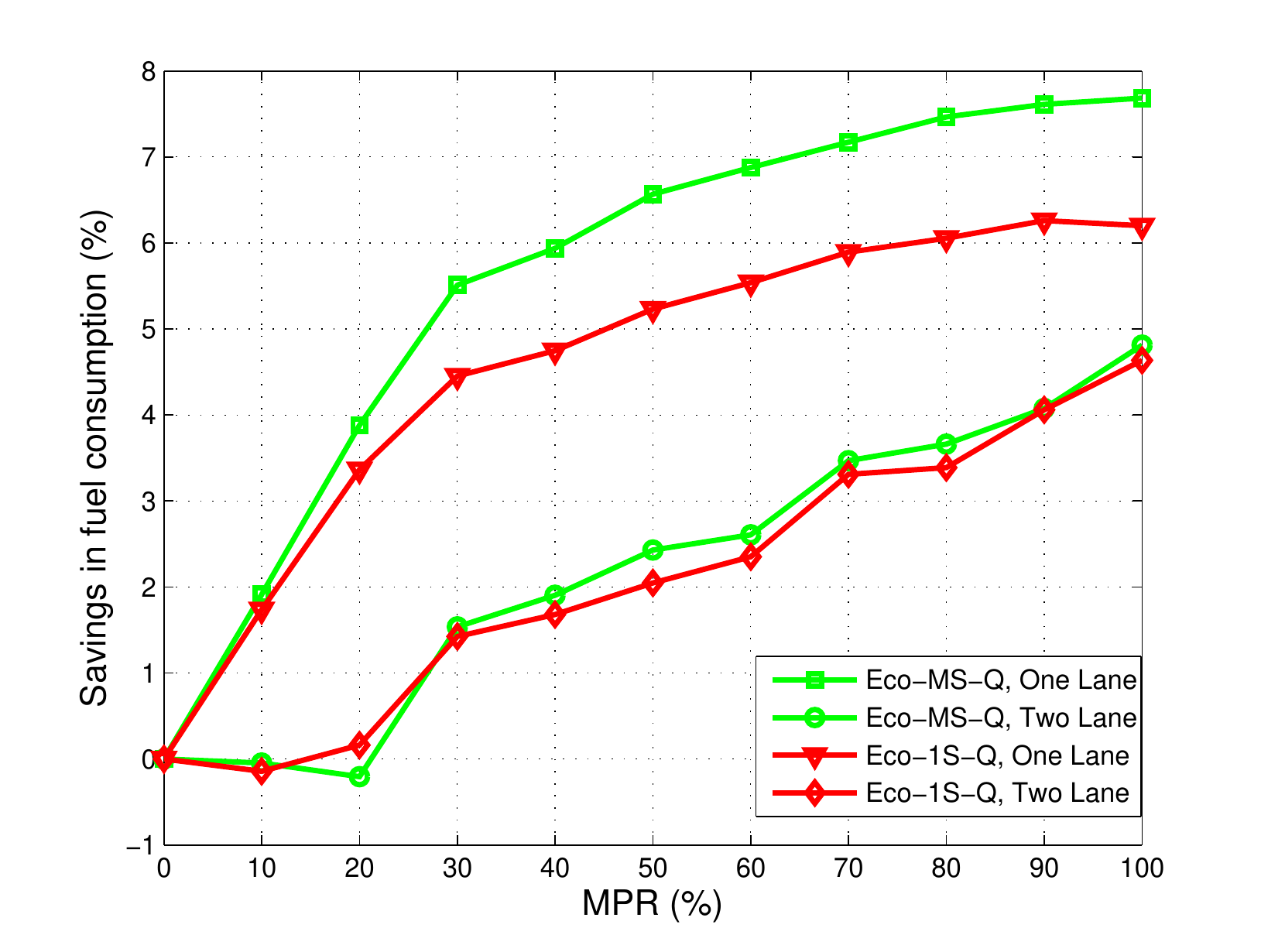}}
\caption{Eco-driving at four intersections: (a) configuration, (b)
savings in fuel consumption.} \label{ms4sig} \ec\efg

\subsection{Testing Eco-MS-Q Algorithm on Grid Network}
In this subsection, the Eco-MS-Q system is implemented and tested on
a grid network composed of 16 signalized intersections (see
\reff{msgrid}(a)). The scale of the network is 2.5 km $\times$ 2.5
km, and the lengths of all links are 500 meters long. The road speed
limits, saturation flow rates, and jam densities are all the same at
80 km/h, 2000 veh/h/lane, and 160 veh/km/lane, respectively. In
addition, all links have two lanes, except the horizontal links
between the nodes 2 and 6 and the vertical links between the nodes
11 and 15. Also, there are one additional left-turn lanes for the
links approaching the intersections. Consequently, different
intersection geometries, including the crossings of two major roads,
two minor roads, and one major and one minor road, are simulated.
For the SPaT plan, the cycle lengths are set between 80 seconds and
100 seconds, and the green-to-cycle ratio varies from 0.2 to 0.3 for
all phases. Furthermore, vehicles are loaded for all OD pairs
combinations.\footnote{The SPaT plan in this scenario is different from the first two scenarios. A more realistic setting is applied here for the grid network to release traffic from all OD pairs.} 

The network is simulated for two hours to evaluate the performance of the proposed Eco-MS-Q system. \reff{msgrid}(b) illustrates the fuel consumption savings of connected (green-dash line), non-connected (blue-dotted line), and all vehicles (red solid line) for different MPRs. Clearly, with higher MPRs, the fuel consumption savings of all three types of vehicles are larger. When the MPR is 100\%, the fuel consumption is reduced by up to 15\%. In addition, the results demonstrate that the fuel wasted ahead of the signals are higher. While with the proposed system, the connected vehicles can be more fuel efficient. Consequently, even with two-lane roads, the passing of non-connected vehicles can only reduce the system performance marginally, and the overall network performance can still be improved.  Moreover, for the connected vehicles, even for very low MPRs, they can gain very high fuel efficiency. At higher MPRs, the improvement is marginal. Furthermore, with the control, the connected vehicles always outperform non-connected ones. On the other hand, the fuel consumption of the non-connected vehicles can also be reduced as a result of the improved network performance.

This trend in \reff{msgrid}(b) is a little different from the previous examples for the two-lane roads. Several explanations can explain these observed differences. First, there exist some minor roads and left turning traffic at the intersections, which can only travel in a single lane. Consequently, connected vehicles traveling on the minor roads or making left turns, always incur positive savings that impact both connected and non-connected vehicles. Second, due to the complexity of the network, many vehicles travel on routes with left turns and/or minor roads. So, even though they consume more fuel on major roads at lower MPRs, they also consume less fuel for left turning traffic and on minor roads, and the combined effect can potentially result in overall fuel savings. Third, in the grid network, all signals have more than two phases, and the green split of each phase is smaller than that in the first two scenarios. The smaller green split produces longer idling times during the red indication and higher probability of stopping, as well as higher fuel consumption. The proposed algorithm can avoid the negative impact of the smaller green split. Consequently, the fuel savings can be higher than the two scenarios in Section III.B and C. Hence, considering the three factors, the savings can be positive even at low MPRs in the grid network.

\bfg \bc
\subfigure[]{\includegraphics[width=3in,height=2.0in]{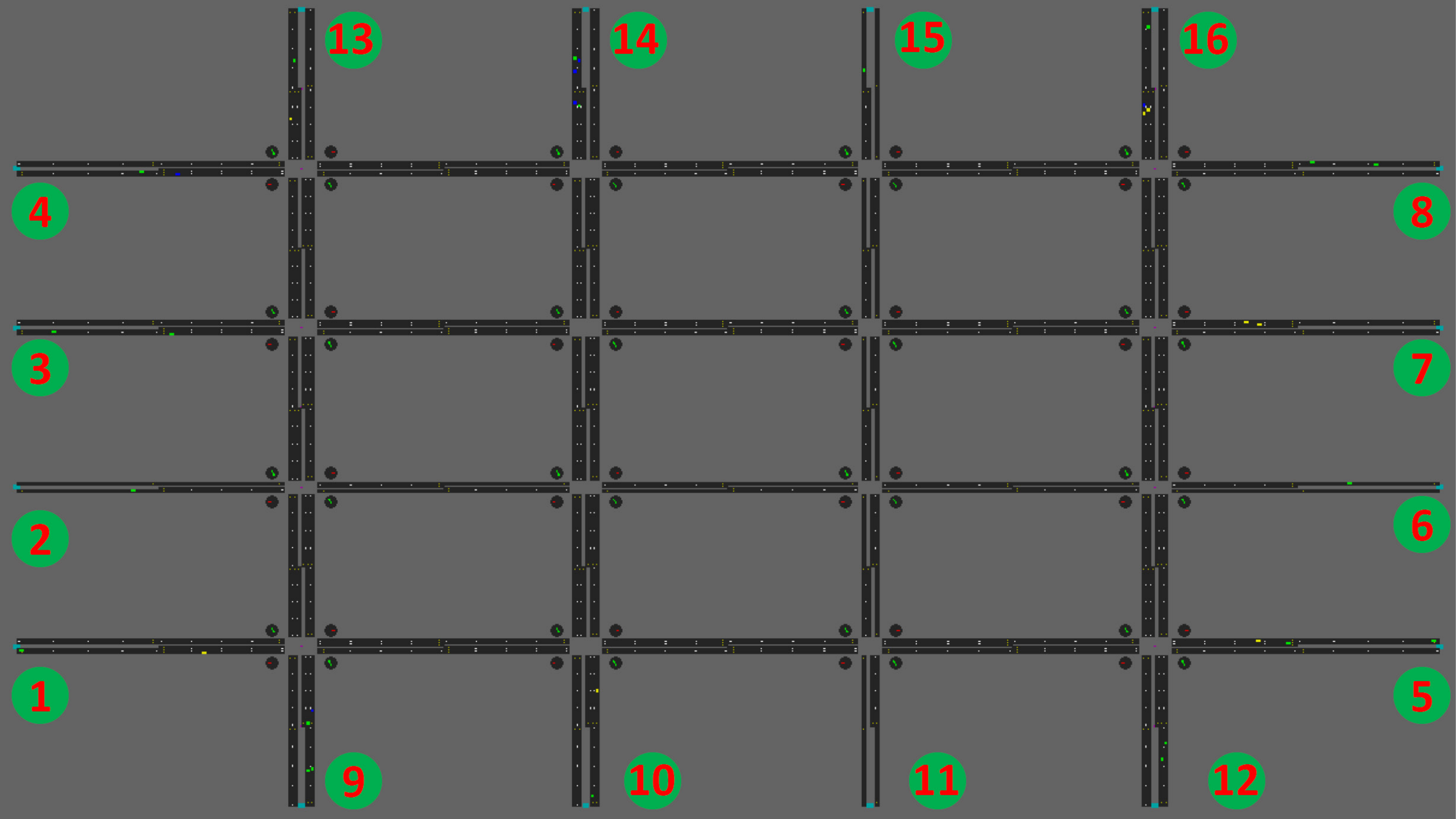}}
\subfigure[]{\includegraphics[width=3.3in,height=2.3in]{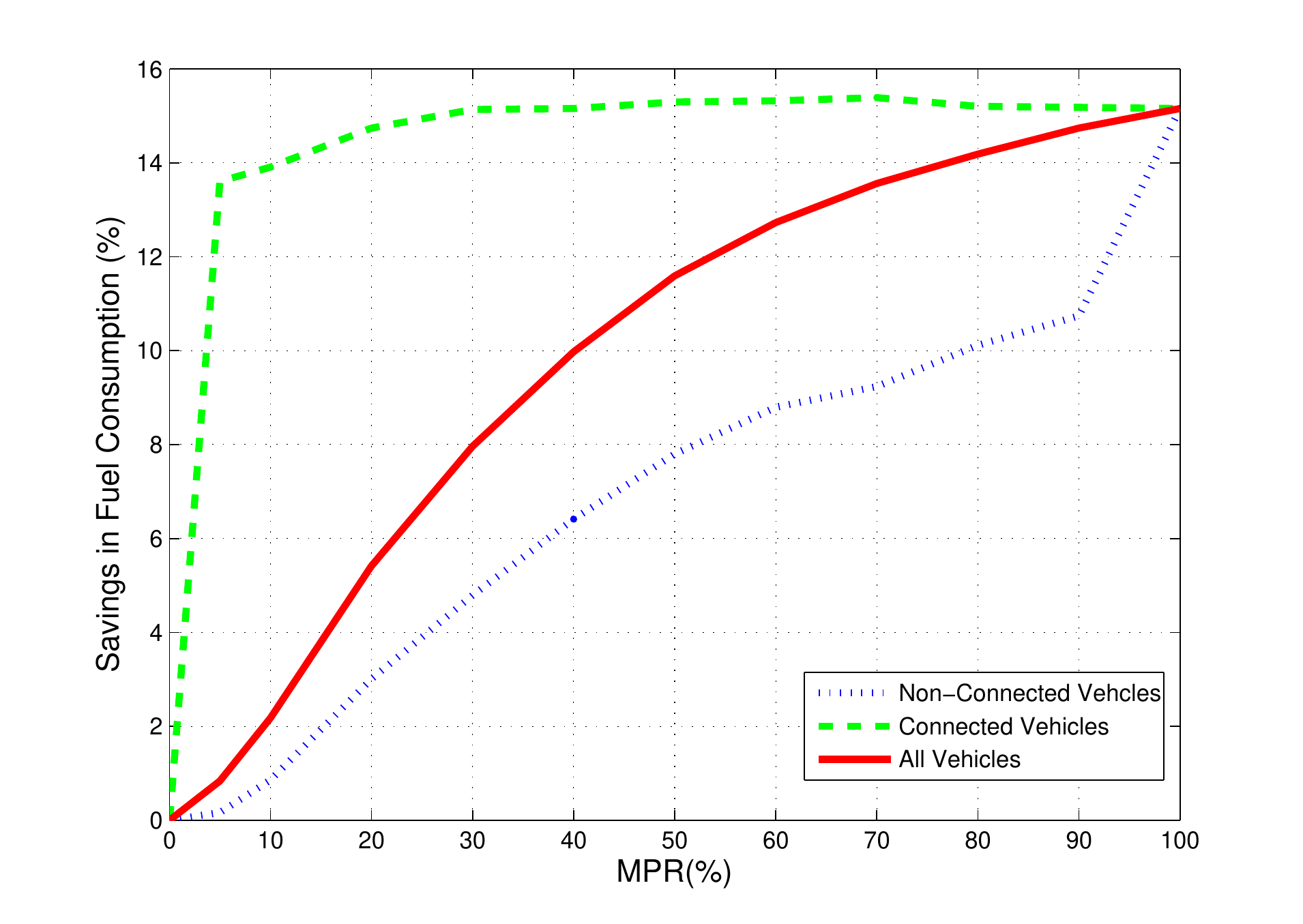}}
\caption{Eco-MS-Q at a grid network (a) network geometry, (b)
savings in fuel consumption.} \label{msgrid} \ec\efg

\section{Sensitivity Analysis}
This section conducts a sensitivity analysis of variables applied in
the proposed eco-driving algorithms, including the traffic demand
levels, phase splits, offsets between two consecutive signals, and
intersection spacings. In addition, the impact of over-saturated
traffic on the algorithm is assessed. Moreover, the Eco-1S-Q and the
Eco-MS-Q algorithms are compared to identify the advantage of the
multiple intersection control. To complete the analysis, the network
in \reff{simsig} with two intersections is simulated, and link
properties, including the road speed limits, the road saturation
flow rates, the jam densities, and the density at capacity, are the
same as that in section III.A.

\subsection{Sensitivity to Demand Levels}
On arterial roads, vehicle demands are directly related to queue
lengths ahead of traffic signals and the number of the equipped
vehicles in the network, and they play an important role in the
performance of the intersections and assessing the the benefits of
the eco-driving algorithms. In this section, we examine the fuel
efficiency of the Eco-MS-Q and Eco-1S-Q algorithms for different
demand levels. The signal settings of the two intersections are the
same as those presented earlier in section III.A. \footnote{The
75-second offset is also applied to observed obvious different
between the single and multiple intersection control strategies.} In
addition, we assume that all vehicles are equipped (i.e., the MPR is
100\%), and the demand varies from $100$ to $700$ veh/h/lane for both
algorithms.

\reff{msdemand} illustrates the fuel consumption savings of both
algorithms as a function of the demand. The results show that under
the given settings of the signal plans and the offset, positive
savings in fuel consumption can be observed for all demand levels.
In addition, for the Eco-MS-Q algorithm, the demand at $400$ veh/h/lane
results in the best savings for the whole network, about 13.5\%.
Demands from $400$ veh/h/lane to $700$ veh/h/lane result in savings of 7\% for
the Eco-1S-Q algorithm. The savings are a result of the increase in
the number of equipped vehicles in the network. However, this
simulation only considers demands below the saturated flow (800
veh/h/lane). The implementation of the algorithm in the over-saturated
network will be different, and the details will be analyzed in
section IV.E.

\bfg \bc
\includegraphics[width=3in,height=2.2in]{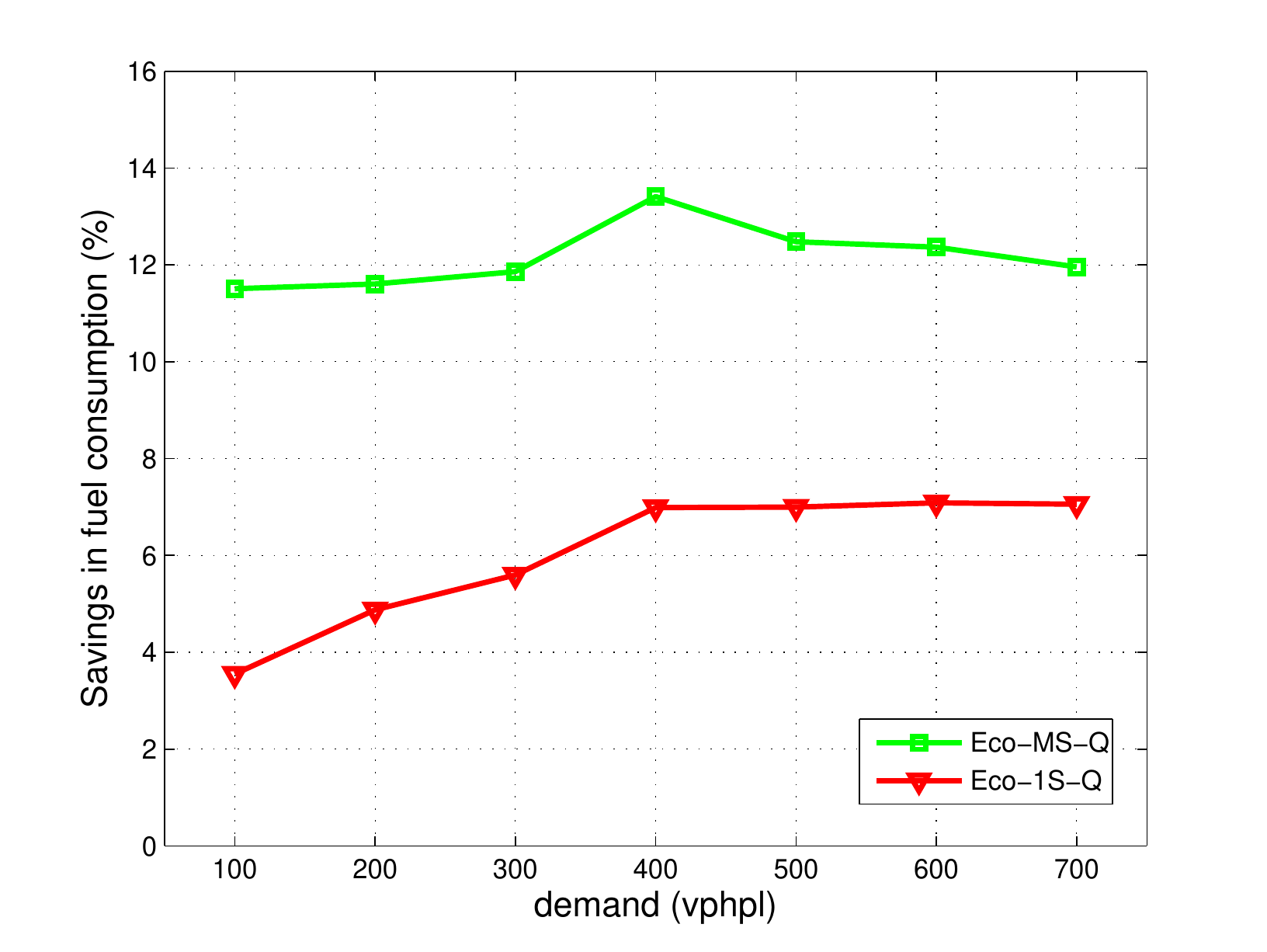}
\caption{Savings in fuel consumption for different traffic demand levels.} \label{msdemand} \ec\efg

\subsection{Sensitivity to Phase Splits}
In this study, both Eco-MS-Q and Eco-1S-Q algorithms utilized the
SPaT information to compute the optimal trajectories for the
equipped vehicles, which indicates the phase splits are highly
related to the effectiveness of the algorithms. In this subsection,
the impact of the phase splits to the overall network performance is
investigated. The simulation settings are the same as those used in
the example in section III.A. The demand is constant as $600$
veh/h/lane,and the phase split (i.e., the ratio of the total duration of
the green and amber indicator to the cycle length) ranges from 35\%
to 75\% for the major road (through traffic from west to east)
respect to the total cycle length of 120 seconds.

\reff{msphsplit} illustrates the fuel consumption savings of both
algorithms as a function of the phase split. The figure demonstrates
that with longer phase lengths, the fuel savings decrease. With a
35\% green split for the major road, the savings reach up to 13.8\%
for the Eco-MS-Q and 7.2\% for Eco-1S-Q. These results are
intuitively correct given that with longer phase lengths, the
equipped vehicles have less chance of stopping at the traffic
signals, resulting in lower fuel consumption savings, as less
vehicles have to stop. Consequently, it only needs to control the
behaviors of less vehicles for longer phase lengths. In addition,
the benefits of the proposed algorithms come from the control of the
stopped vehicles. In that sense, the savings of fuel consumption
will be smaller.

\bfg \bc
\includegraphics[width=3in,height=2.2in]{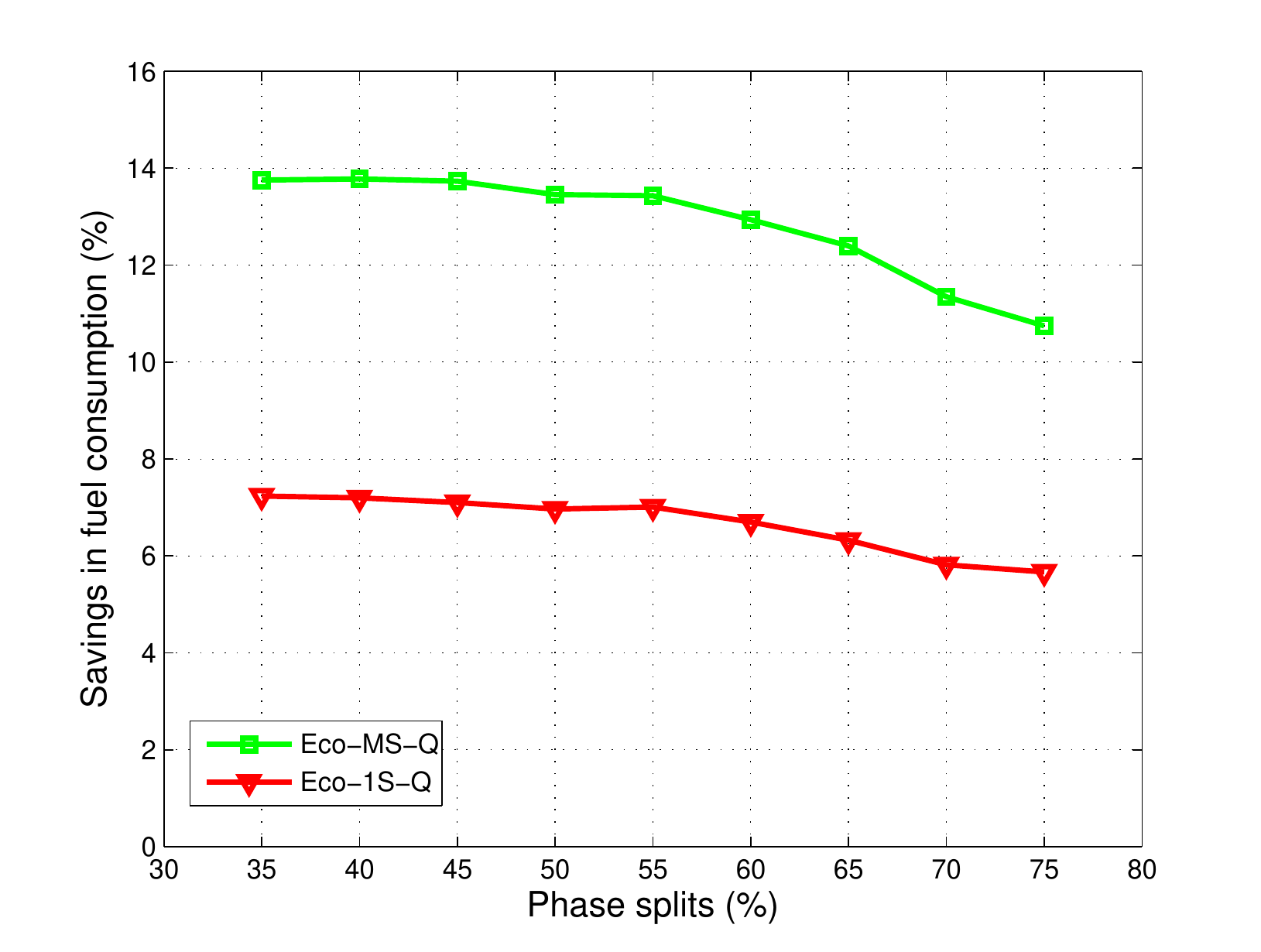}
\caption{Savings in fuel consumption for different phase splits.}
\label{msphsplit} \ec\efg

\subsection{Sensitivity to Traffic Signal Offsets}
The offset is important to coordinate multiple intersections and to
improve the network performance. Furthermore, the fluctuations in
vehicle movements through multiple intersections are directly
related to the offset. This section investigates the impact of
offsets on the performance of the proposed algorithms. The
simulation settings are still the same to the example in section
IV.B, except that the phase split is constant at 55\%, and the
second signal offset with respect to the first varies from $0$ to
$120$ seconds. In this simulation, the distance between the two
intersections is 1000 meters, which can be traveled by equipped
vehicles within 45 seconds at the free-flow speed, i.e., the optimal
offset of the second signal is about 45-50 seconds (with
consideration of lost time).

\reff{msoffset} illustrates the fuel consumption savings of both
Eco-MS-Q and Eco-1S-Q algorithms as a function of the offset.
Results indicate that when the offset is closer to the optimal
value, the fuel savings obtained from both algorithms decrease. At
the optimal offset, the Eco-MS-Q provides the lowest saving of
2.8\%, and the Eco-1S-Q 2.5\%. The highest savings of fuel
consumption can be observed at 13.0\% with a 100-second offset for
Eco-MS-Q, and 7.3\% with a 65-second offset for Eco-1S-Q. These
results are expected, as the savings of the algorithms are observed
if the equipped vehicles have to stop at both intersections. At the
optimal offset, most vehicles only need to stop at the first traffic
signal, which results in the least savings for both algorithms.

\bfg \bc
\includegraphics[width=3in,height=2.2in]{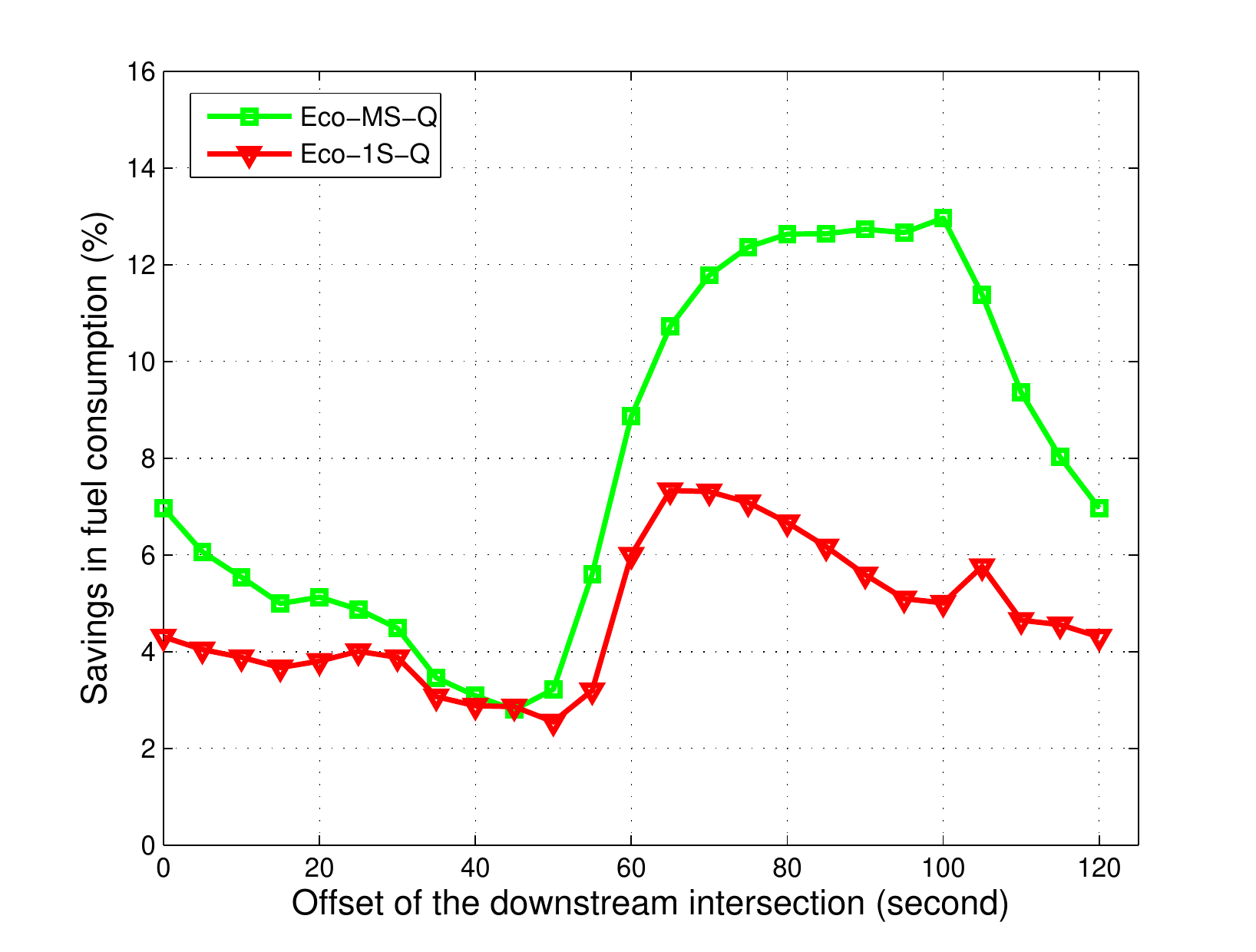}
\caption{Savings in fuel consumption for different offsets.}
\label{msoffset} \ec\efg

\subsection{Sensitivity to the Intersection Spacing}
Distance between intersections is another variable affecting the
benefits of the proposed algorithm. In this section, the impact of
the distance between the intersections is evaluated. The simulation
settings are the same at those presented in section IV.C, except
that the offset is constant as $75$ seconds, and the distance
between the two intersections ranges between $200$ and $1000$
meters.

\reff{msdistance} shows the fuel consumption savings for both
Eco-MS-Q and Eco-1S-Q algorithms as a function of the intersection
spacing. Results indicate that for the given signal plans and demand
level, $700$ meters is the optimal distance between intersections
for both algorithms. The Eco-MS-Q algorithm provides fuel
consumption savings of 13.1\% for the 700-meters spacing, and the
Eco-1S-Q algorithm provides 7.2\% savings.

The pattern from the Eco-MS-Q algorithm is determined by the
following two factors. (1) With a longer intersection spacing,
equipped vehicles can be controlled for a longer time, allowing the
algorithms to provide more fuel-efficient trajectories. However, (2)
the longer distance makes the prediction of the queue lengths and
queue dissipation times at the downstream intersection less
accurate, which reduces the effectiveness of the algorithm. Hence,
there exits an optimal value for the distance when the Eco-MS-Q
algorithm is applied. In addition, with the Eco-1S-Q algorithm, the
two intersections are controlled independently. When the distance is
large enough, the two intersections can be considered to be
isolated. Hence, the benefit from the single intersection control
remains constant.

\bfg \bc
\includegraphics[width=3in,height=2.2in]{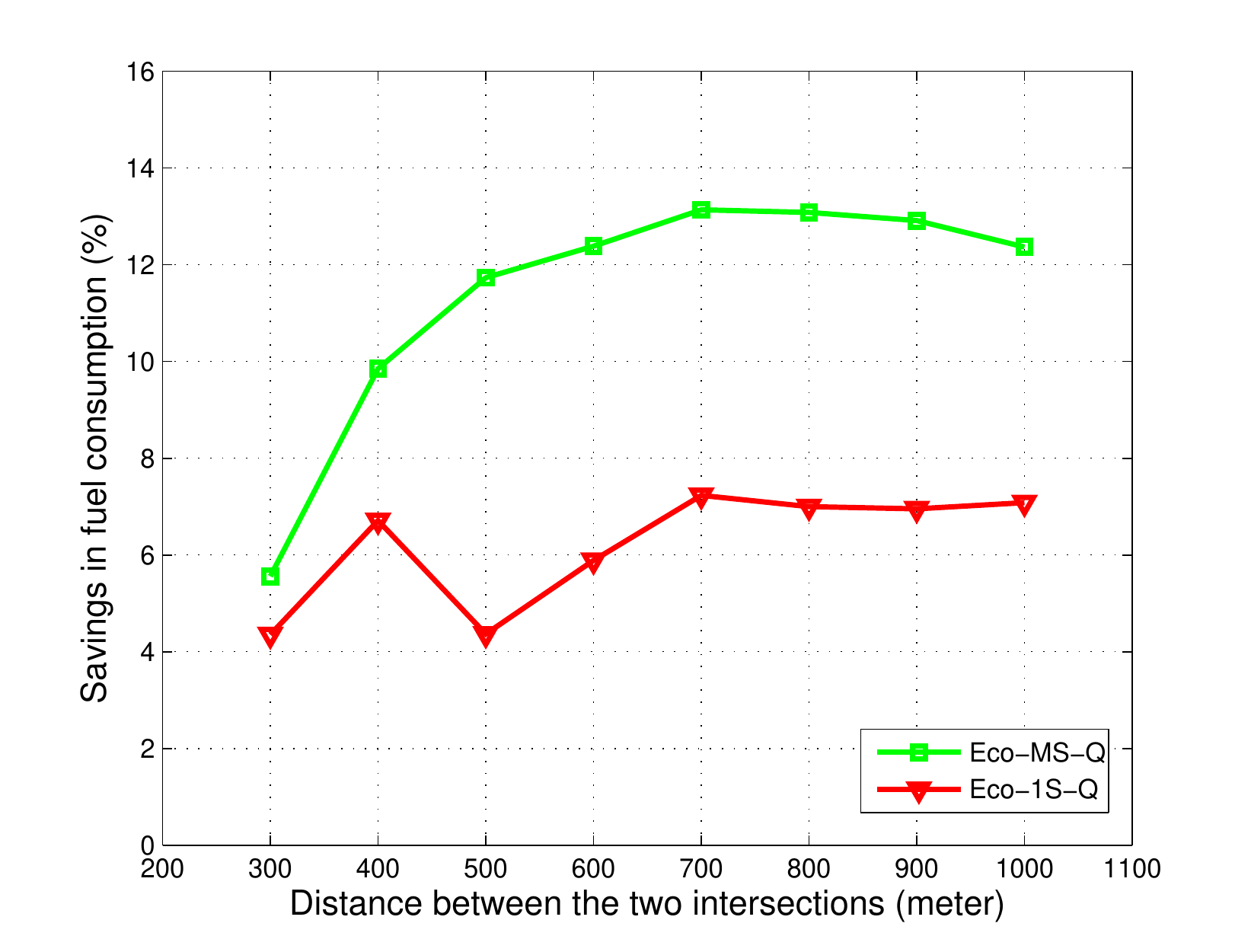}
\caption{Savings in fuel consumption for different intersection spacings.} \label{msdistance} \ec\efg

\subsection{Over-Saturated Demands}
In the development of the Eco-MS-Q algorithms, one critical
assumption is that the network is not over-saturated, and vehicle
queues can be released in a single cycle. Once the network is
over-saturated, rolling queues are generated upstream of the
intersections. Then, the queue estimation method in
\cite{yang2017eco} cannot determine the queue length and the
dissipation time accurately. Consequently, the advisory speed limits
are not optimized adequately. \cite{yang2017eco} showed that the
Eco-1S-Q algorithm was unable to obtain positive savings for
over-saturated demands. In this section, we investigate the impact
of over-saturated demands on the Eco-MS-Q algorithm.

The example in section III.A is applied. The simulation settings are
the same, except that the demand increases to $1000$ veh/h/lane, which is
grater than the capacity of the controlled segment. \reff{overs}
compares the trajectories of all vehicles before and after applying
the Eco-MS-Q algorithm. In \reff{overs}(a), most vehicles experience
stops at both intersections, and generally the queues at the first
intersection cannot be released in one cycle. \reff{overs}(b) shows
the trajectories with 10\% equipped vehicles. As shown within the
black box in the figure, the algorithm fails to provide the optimum
speed limits for the equipped vehicles to proceed through the
intersection without stops. The rolling queues caused by the
over-saturated demand levels generate traffic fluctuations and
complete stops for the equipped vehicles, reducing the benefits of
the algorithm dramatically. However, as the inflow to the second
intersection is gated by the first one, over-saturation is averted,
and fuel consumption savings can still be observed at the second
intersection using the proposed algorithm. In the simulation, a 10\%
MPR actually reduces fuel consumption by about 2.7\%. This implies
that compared with Eco-1S-Q, the Eco-MS-Q algorithm is more
efficient at providing fuel consumption savings under over-saturated
demands.

\bfg \bc
\subfigure[]{\includegraphics[width=3.3in,height=1.2in]{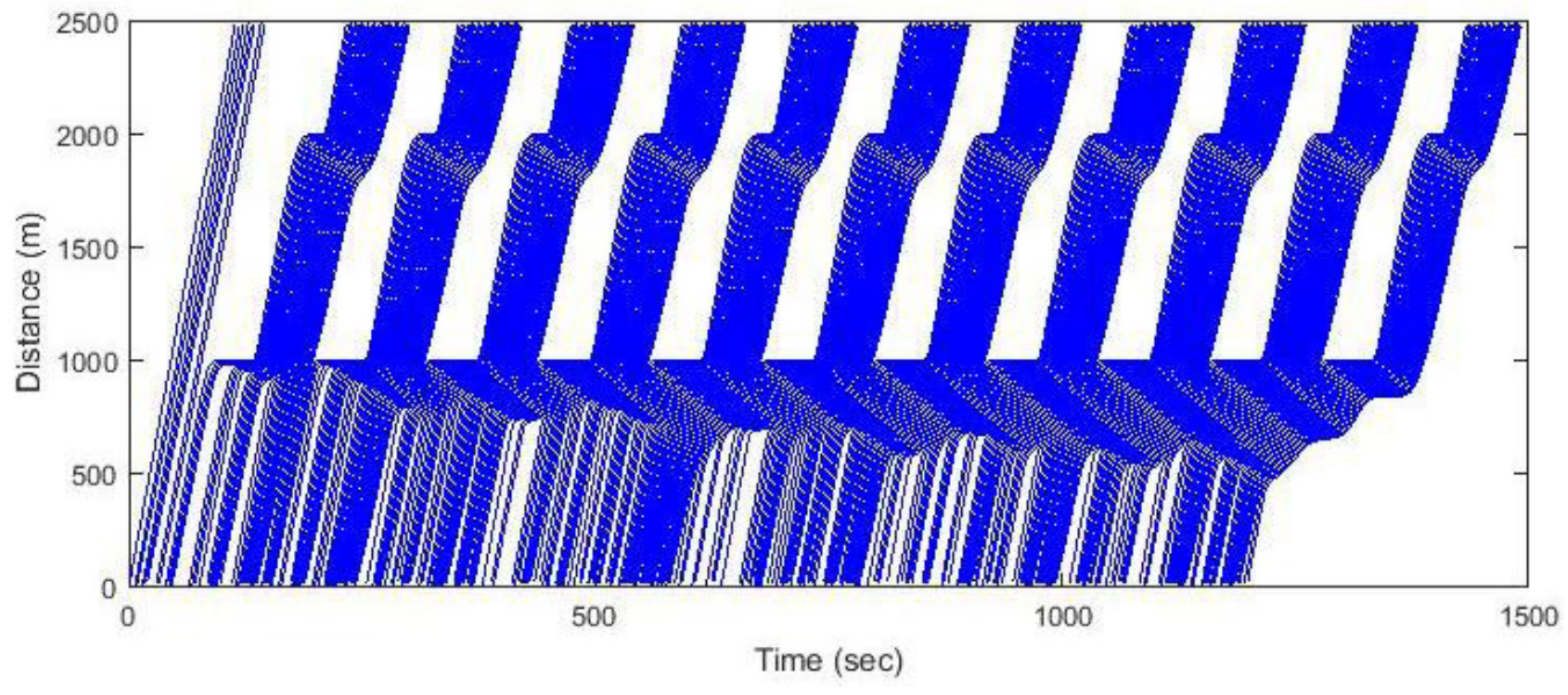}}
\subfigure[]{\includegraphics[width=3.3in,height=1.2in]{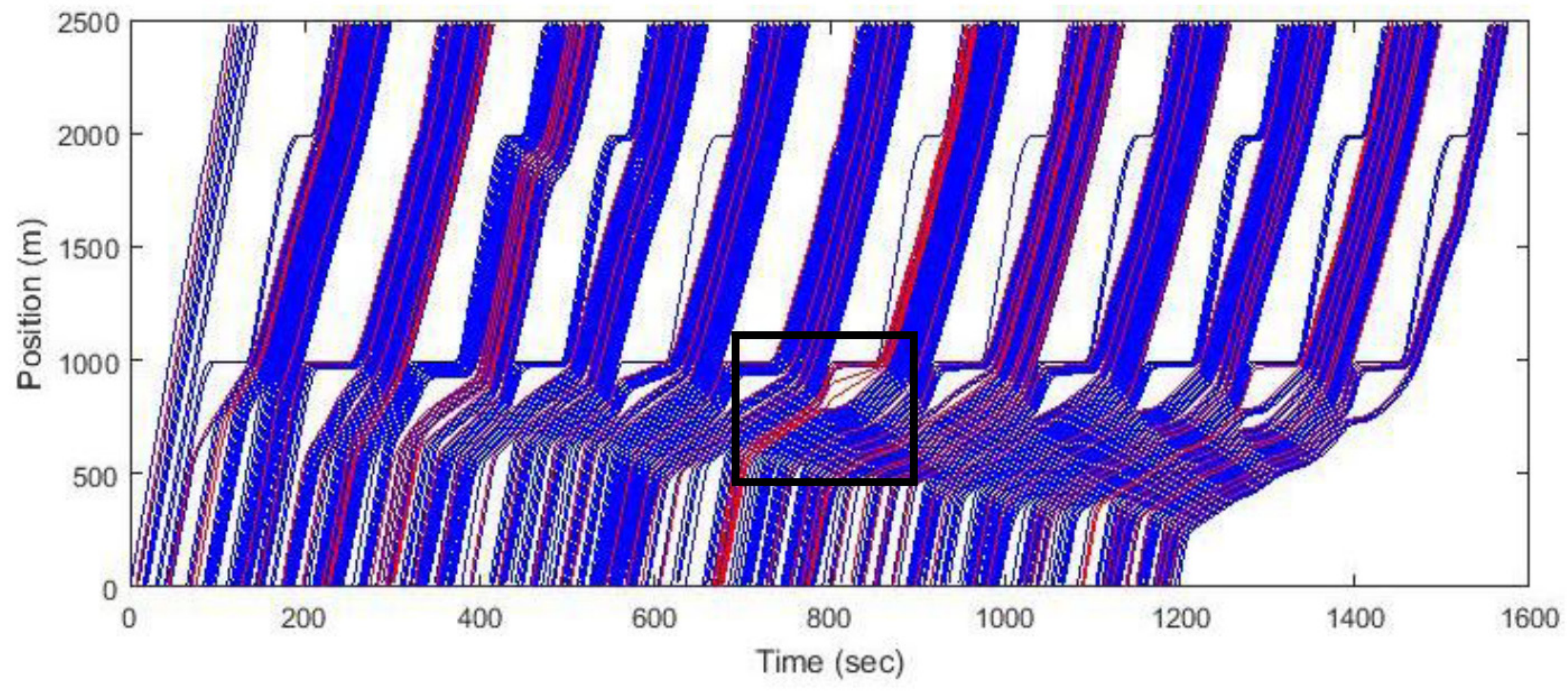}}
\caption{Vehicle trajectories for over-saturated demands: (a) No
control, (b) Eco-MS-Q.} \label{overs} \ec\efg

\section{Conclusions}
The research presented in this paper developed an Eco-MS-Q algorithm
to minimize vehicle fuel consumption levels traveling through
multiple signalized intersections. The algorithm utilizes SPaT
information collected via V2I communications and predictions of
queues to estimate the optimal vehicle trajectories that are provide
and update desired speeds within an ACC system. The algorithm allows
CVs to decelerate/accelerate to a constant cruising speed while
traversing intersections so as to reduce their fuel consumption
levels. The INTEGRATION simulation of the single-lane intersections
demonstrated that fuel consumption savings were greater at higher
MPRs. The reductions in fuel consumption reached 7\% for Eco-MS-Q,
compared with 4.2\% for Eco-1S-Q, at 100\% MPR. In addition, taking
the vehicle queue into consideration, the Eco-MS-Q algorithm always
performed better than Eco-MS-O. In the case of two-lane approaches,
due to lane-changing and passing maneuvers, the proposed algorithm
increased the total fuel consumption levels when CV MPRs were less
than 30\%. Once the MPRs were larger than 30\%, positive savings
could be observed. In addition, the Eco-MS-Q algorithm was
implemented and tested on a four-signal arterial producing fuel
consumption savings as high as 7.7\% for single-lane roads, and
4.8\% for two-lane roads. The algorithm could also effectively
increase vehicle energy efficiency in a grid network composed of 16
signalized intersections producing benefits for both CVs and
non-CVs. These benefits were observed for all CV MPRs even for
multiple lane roadways. CVs always had better performance than
non-CVs.

The study also included a comprehensive sensitivity analysis of
traffic demand levels, phase splits, offsets, and signalized
intersection spacings. The analysis demonstrated that for the given
offset of 75 seconds, a 50/50 phase split, and traffic signal
spacing  of 1000m, loading vehicles at $700$ veh/h/lane resulted in
the highest fuel consumption savings, of 13.5\%. Furthermore, given
the offset, the demand and the distance between intersections, with
a larger percentage of green allocated to the approach, the savings
from the proposed algorithm were smaller. In addition, when the
offset was closer to the optimal offset, fuel consumption savings
were smaller. Furthermore, the optimal traffic signal spacing exists
to maximize fuel consumption savings.

Currently, the proposed algorithm does not operate efficiently for
over-saturated intersections due to the impact of rolling queues. In
the future, we plan to apply V2V communication to collect more
information from individual vehicles and develop a more accurate
queue prediction model. We could also introduce a speed
harmonization algorithm \cite{yang2016development} to restrict
traffic entering the intersections to maintain under-saturated
traffic conditions at all times. Moreover, we will extend the logic
to actuated and adaptive traffic signals. We also plan to conduct a
further sensitivity analysis to study the impact of speed limits,
number of lanes, and road grades on the system performance.
Furthermore, the impacts of wireless communications, such as
communication delay, packet loss, and transmission range, will be
analyzed.

\bibliographystyle{IEEEtran}
\bibliography{es_reference}

\begin{IEEEbiography}[{\includegraphics[width=0.9in,height=1.4in,clip,keepaspectratio]{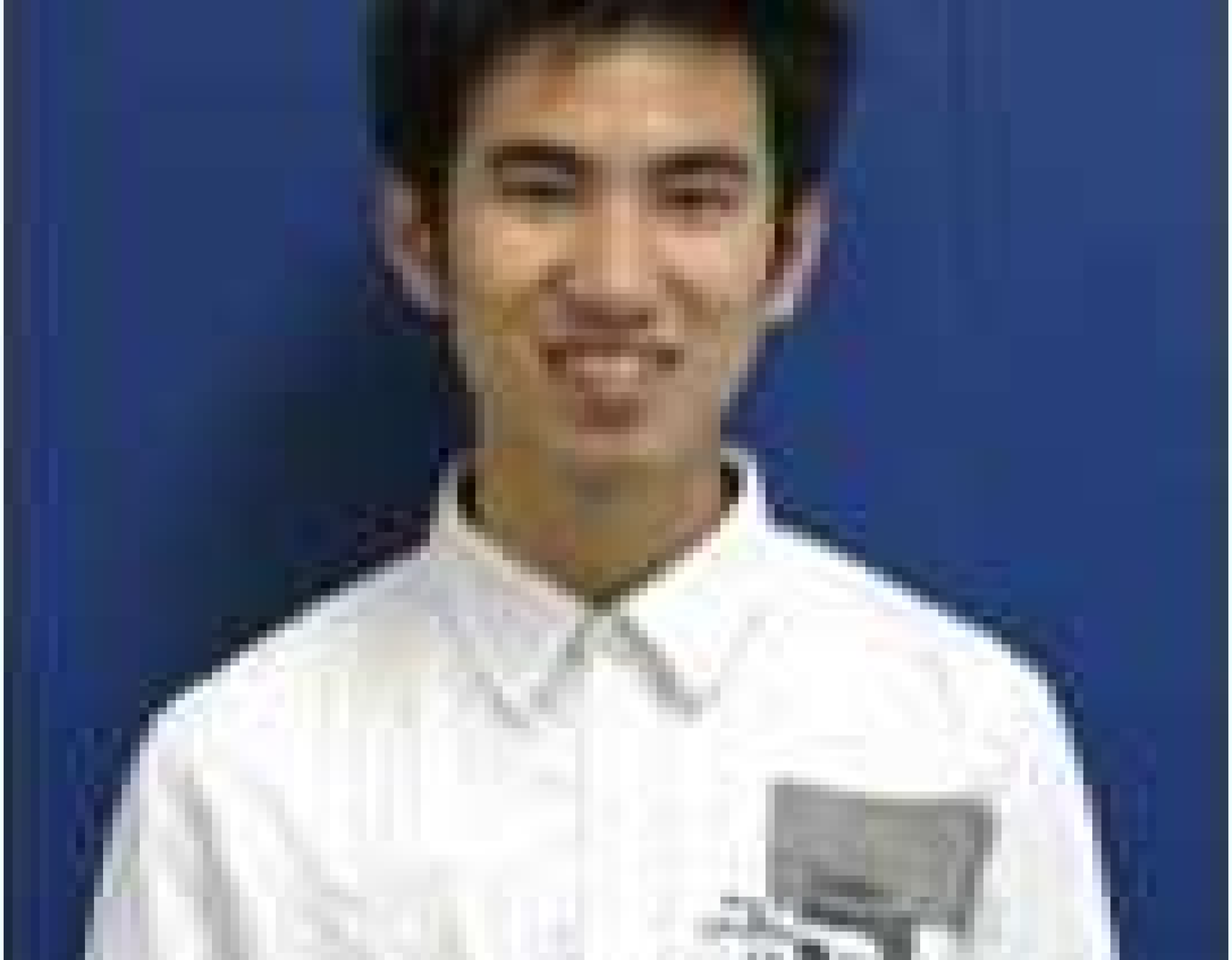}}]{Hao Yang}
received B.S. degree in Automation from University of Science and
Technology of China, China, in 2008. He received M.S. and Ph.D.
degrees in Civil and Environmental Engineering from University of
California (UC), Irvine, USA. He worked as a Postdoctoral Associate
in Virginia Tech Transportation Institute, Blacksburg, VA USA. He
worked as a Visiting Assistant Professor in Lamar University,
Beaumont, TX USA. Currently, he works at Toyota
InfoTechnology Center, Mountain View, CA USA. Dr. Yang's research
interests include intelligent transportation systems, connected and
autonomous vehicles, environmental sustainability, traffic
simulations, transportation data analysis.
\end{IEEEbiography}

\begin{IEEEbiography}[{\includegraphics[width=0.9in,height=1.4in,clip,keepaspectratio]{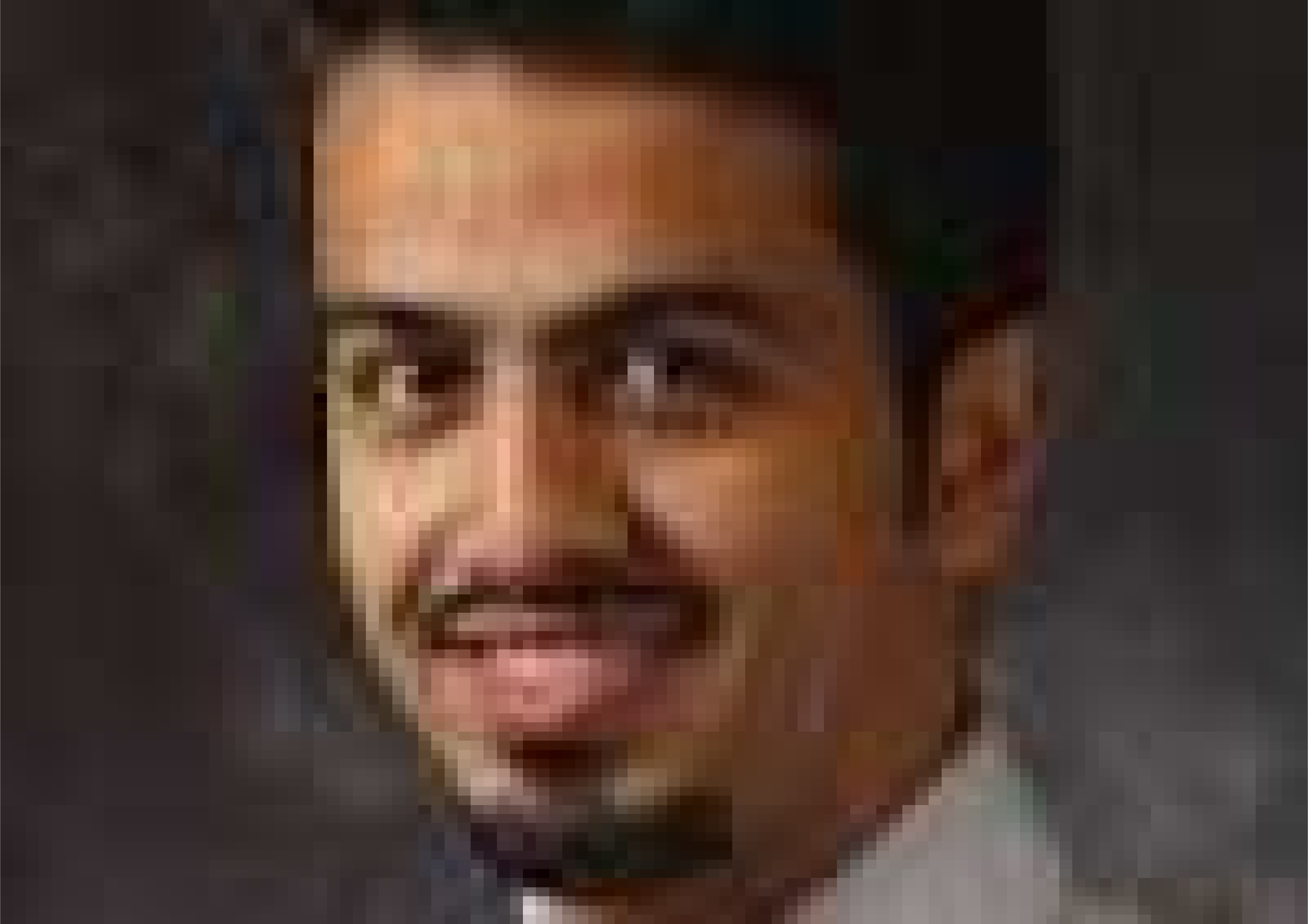}}]{Fawaz Almutari}
received B.S. degree in Civil Engineering from Northern Arizona
University, Flagstaff, AZ, USA, in 2014. He received M.S. degree in
Civil and Environmental Engineering from Virginia Polytechnic
Institute and State University, Blacksburg, VA, USA, in 2016.
Currently, he is continuing pursuing Ph.D. degree in Virginia Tech.
His research interests include connected and autonomous vehicles,
environmental sustainability, and traffic signal control.
\end{IEEEbiography}

\begin{IEEEbiography}[{\includegraphics[width=0.9in,height=1.4in,clip,keepaspectratio]{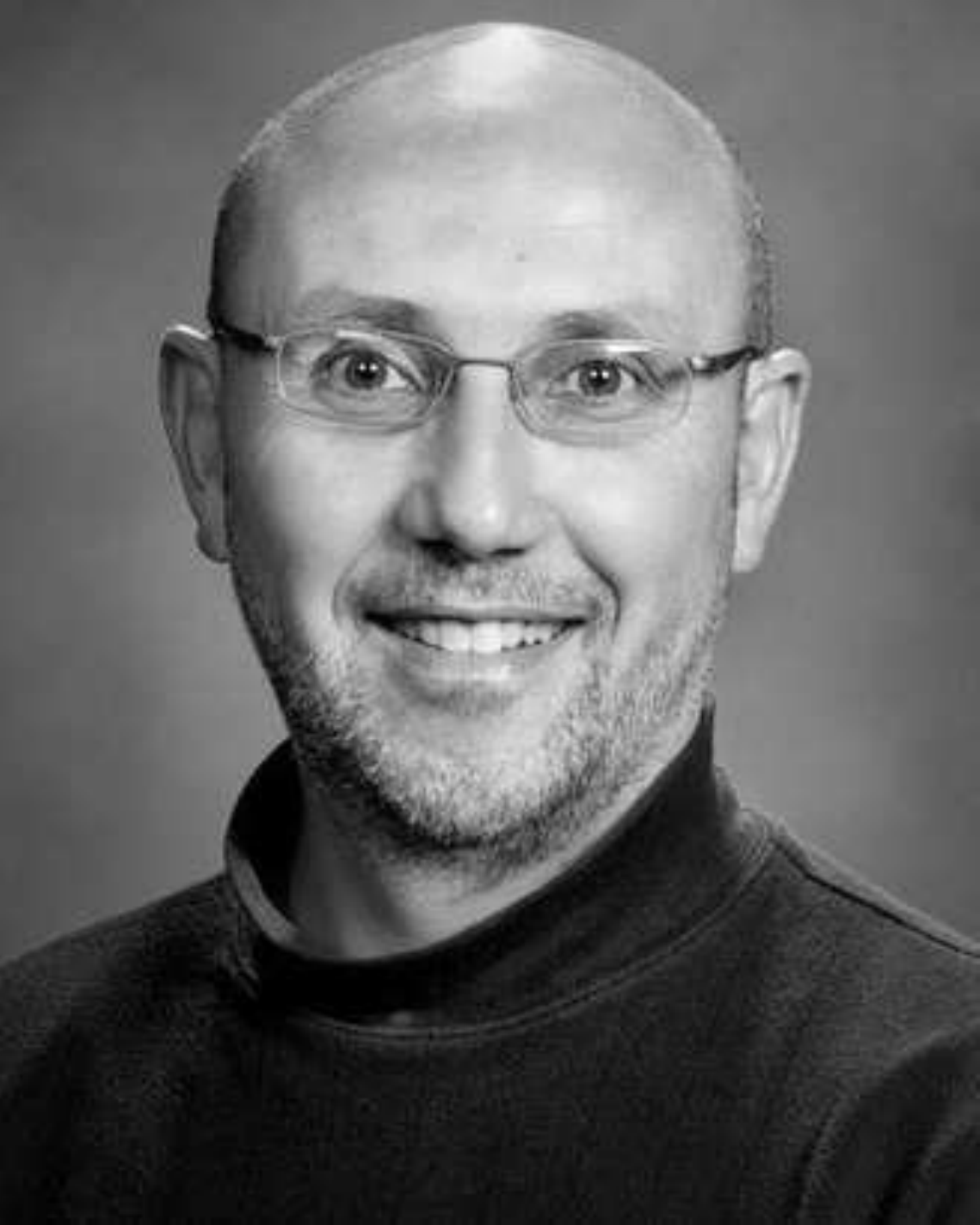}}]{Hesham Rakha}
(M’04, SM’18, F’20) received his B.Sc. (with honors) in civil engineering from Cairo University, Cairo, Egypt, in 1987 and his M.Sc. and Ph.D. in civil and environmental engineering from Queen’s University, Kingston, ON, Canada, in 1990 and 1993, respectively. Dr. Rakha’s research focuses on large-scale transportation system optimization, modeling and assessment. Specifically, Dr. Rakha and his team have expanded the domain of knowledge (in traveler and driver behavior modeling) and developed a suite of multi-modal agent-based transportation modeling tools, including the INTEGRATION microscopic traffic simulation software. This software was used to evaluate the first dynamic route guidance system, TravTek in Orlando, Florida; to model the Greater Salt Lake City area in preparation for the 2002 Winter Olympic Games; to model sections of Beijing in preparation for the 2008 Summer Olympic Games; to optimize and evaluate the performance of alternative traveler incentive strategies to reduce network-wide energy consumption in the Greater Los Angeles area; and to develop and test an Eco-Cooperative Automated Control (Eco-CAC) system. Finally, Dr. Rakha and his team have developed various vehicle energy and fuel consumption models that are used world-wide to assess the energy and environmental impacts of ITS applications and emerging Connected Automated Vehicle (CAV) systems. The models include the VT-Micro, VT-Meso, the Virginia Tech Comprehensive Fuel consumption Model (VT-CPFM), the VT-CPEM, and the VT-CPHEM models.
\end{IEEEbiography}

\end{document}